\documentclass{aa}
\usepackage{graphicx}
\usepackage{txfonts}

\begin{document}

\title{Magnetic field evolution in interacting galaxies}

\author{
R. T. Drzazga \and K. T. Chy\.zy
 \and W. Jurusik
 \and K. Wi\'orkiewicz
}
\institute{Astronomical Observatory, Jagiellonian University, ul. Orla 171, 30-244 Krak\'ow, Poland
}
\offprints{R. Drzazga}
\mail{drzazga@oa.uj.edu.pl}

\date{Received date/ Accepted date}
\titlerunning{Magnetic field evolution in interacting galaxies}
\authorrunning{R. Drzazga et al.}

\abstract
%heading
{}
%aims
{Violent gravitational interactions can change the morphologies of galaxies
and, by means of merging, transform them into elliptical galaxies. We aim to 
investigate how they affect the evolution of galactic magnetic fields.
}
%methods
{We selected 16 systems of interacting galaxies with available VLA archive radio data
at 4.86 and 1.4\,GHz and compared their radio emission and estimated magnetic field 
strengths with their star-forming activity, far-infrared emission, and the stage 
of tidal interaction.
}
%results
{
The estimated mean of total magnetic field strength for our sample of interacting galaxies 
is $14\pm 5\,\mu$G, which is larger than for the non-interacting objects. The field regularity 
(of $0.27\pm0.09$) is lower than in typical spirals and indicates enhanced production
of random magnetic fields in the interacting objects.
We find a general evolution of magnetic fields:
for weak interactions the strength of magnetic field is almost constant
($10-15\,\mu$G) as interaction advances, then it increases up to $2\times$,
peaks at the nuclear coalescence ($25\,\mu$G),
and decreases again, down to $5-6\,\mu$G, for the post-merger remnants.
The main production of magnetic fields in colliding
galaxies thus terminates somewhere close to the nuclear coalescence, after which
magnetic field diffuses.
The magnetic field strength for whole galaxies is weakly affected by the star formation rate (SFR),
while the dependence is higher for galactic centres. 
We show that the morphological distortions visible in the radio total and 
polarized emission do not
depend statistically on the global or local SFRs, while they do increase
(especially in the polarization) with the advance of interaction.
The constructed radio-far-infrared relations for interacting and non-interacting 
galaxies display a similar balance between the generation of 
cosmic rays, magnetic fields, and the production of the thermal 
energy and dust radiation.
}
%conclusions
{
The regular magnetic fields are much more sensitive to morphological distortions
induced by tidal interactions than are the random fields. As a result the polarized
emission could be yet another indicator of an ongoing merging process.
The found evolution of magnetic field with advancing interaction 
would definitely imply a stronger effect of magnetic fields on the galaxy surroundings in the 
earlier cosmological epochs. 
The process of strong gravitational interactions can 
efficiently magnetize the merger's surroundings, having a similar magnetizing effect 
on intergalactic medium as supernova explosions or galactic winds. 
If interacting galaxies generate some ultra-high energy cosmic rays 
(UHECRs), the disk or magnetized outflows can deflect 
them (up to $23\degr$), and make an association of the observed UHECRs with the sites of their origin very uncertain.
}
\keywords{Galaxies: general -- galaxies:  magnetic fields -- galaxies: interactions -- radio continuum: galaxies}

\maketitle

\section{Introduction}
\label{s:intro}

Both observational evidence and theoretical considerations
show that galaxy interactions are the most important processes during the formation and
evolution of galaxies (Struck \cite{struck99}). This is the case both locally in the nearby
universe and, in particular, cosmologically at high redshifts,
where galaxy encounters are expected to be more frequent.
During such encounters, direct collisions between stars are highly unlikely, while 
collisions between the interstellar medium (ISM) of the two 
disks are inevitable and may trigger star formation.
The interaction-induced starbursts, nuclear activity,
feedback, gas infalls, and outflows, all play critical roles in
shaping ISM and the surrounding intergalactic medium (IGM)
(Smith et al. \cite{smith10}).

Some low-speed collisions lead to merging of galaxies. Gas-rich mergers 
could reach the phase of a luminous infrared galaxy (LIRG), or even of an 
ultraluminous infrared galaxy (ULIRG) (Sanders \& Mirabel \cite{sanders96}). 
Violent galaxy collisions are also suggested as a source of large-scale shocks and 
one of possible sources of ultra high energy cosmic rays (UHECR; e.g. Bhattacharjee 
\& Sigl \cite{bhattacharjee00}). However, not all 
galaxy interactions will result in a merger, and it is equally important to 
study different types of galaxy interaction to understand their role in galaxy 
evolution. 

There were several systematic attempts to study the effects of 
collisions on structures of galaxies and the properties of different ISM phases.
An early work of Toomre \& Toomre (\cite{toomre72}) showed that numerical 
simulations quite readily reproduce the observed morphology 
of a wide range of interacting galaxies and explain the formation of bars, 
tidal arms, and bridges. Toomre (\cite{toomre77}) compiled a sample 
of 11 ongoing merger systems at various stages of interaction, which became known as 
the {\em Toomre sequence}. The sequence itself and numerical simulations of 
its members explain how elliptical galaxies can be formed by the coalescence of 
spiral galaxies. The Toomre sequence was studied over the years 
in great detail in various wavelength bands in order to understand the evolving properties 
of interacting systems. It was shown that in moving along the sequence from early 
to late-stage mergers, a larger fraction of atomic gas appears outside 
the optical bodies (Hibbard \& Gorkom \cite{hibbard96}). 
In the final stages of nuclear coalescence, there is little \ion{H}{i} within 
the remnant bodies, while the tidal material flows in and is able to 
trigger a nuclear starburst. In accordance with that, Hummel et al. (\cite{hummel90}) 
found that the central sources of interacting galaxies are by a factor of five times 
brighter in the radio wavelengths than those in the isolated spirals. 

The studies of broadband (V, I) and narrowband (H$\alpha$) images 
obtained with the Hubble Space Telescope (HST) by Laine et al. (\cite{laine03}) 
show surprisingly little evidence of any distinct trends in nuclear properties along the
Toomre merger sequence. There is just a rise in the nuclear
luminosity density in the most-evolved members of the
sequence. However, the near-infrared (NIR) observations with the HST (Rossa et al. \cite{rossa07}), 
which are less affected by dust obscuration show a 
distinct trend for the nuclei in the sequence to become more luminous as the merger process advances.
In a different way than the non-interacting objects, the Toomre-sequence galaxies also reveal newly 
formed stars that are more concentrated toward their centres.
Rossa et al. also argue that if left to evolve and fade for several Gyrs, 
the merger remnants would have the $K$-band luminosity profiles of normal elliptical galaxies.

More complicated evolution during the merging process was identified by Brassington et al. 
(\cite{brassington07}) in the X-rays observed with {\it Chandra}.
They studied a sample of nine interacting and post-merger objects. The original Toomre sequence 
was extended in this work to include very old (up to a 3\,Gyr) merger-remnants. 
First, they find that unlike H$\alpha$ and optical 
emission, the X-ray luminosity peaks about 300\,Myr before the nuclear coalescence. 
The subsequent drop in X-ray emission is likely due to large-scale diffuse outflows
coming out of the galactic disks and reducing the hot gas density. 
Second, about 1 Gyr after the coalescence the remnants are X-ray-fainter than 
typical elliptical galaxies, which are known to be generally strong X-ray emitters.
Thirdly, the dynamically older mergers seem to rebuild the hot galactic halo
and rise again the X-ray luminosity. This can be explained by an outflowing wind driven 
into the halo by a declining population of Type Ia supernova (Brassington 
et al. \cite{brassington07}).

Although magnetic field constitute an important ingredient of the ISM,
little is known about how galaxy interactions affect its strength and evolution, and
how they respond to the evolution of other ISM phases presented above . 
The most detailed analysis of magnetic field in disks of interacting galaxies was 
done by Chy\.zy \& Beck (\cite{chyzy04}) for the Antennae galaxies.
They show that the mean total magnetic field strength (about $20\,\mu$G) 
is significantly greater than in isolated grand-design spirals and may be due 
to the interaction-induced star formation. The field regularity (the ratio of 
regular to random field components) appeared to be lower than in undistorted 
galaxies, which can indicate enhanced turbulent motions in the merger system. 
A strong regular component (of $10\,\mu$G) was found to trace gas-shearing motions 
along the tidal tail. When and how the enhanced magnetic field arose and when 
it would disperse in this system are still not known. 

The evolution of magnetic field in merging systems like the Antennae 
may look different from the evolution of the X-ray component revealed by 
Brassington et al. (\cite{brassington07}). 
In general, the early-type galaxies are typically radio weak, and 
strong radio emission is only observed in ellipticals with an AGN activity. 
In such active galaxies, generation of strong magnetic fields, although possible in 
accretion disks, is not connected with any large-scale or small-scale galactic 
dynamo processes, which operate in typical spiral or irregular galaxies 
(Beck et al. \cite{beck96}, Brandenburg \& Subramanian \cite{brandenburg05}, Chy\.zy 
et al. \cite{chyzy11}). The first process requires a strong differential rotation with uniform sign of 
turbulence twisting, the second depends on the ISM turbulence, which likely declines 
in early-type galaxies owing to a weaker star formation. 

The goal of this paper is to investigate the evolution of magnetic field 
during galaxy interactions. As magnetic fields are best studied from 
the synchrotron emission, we restrict our considerations to objects with available radio data. The 
polarized properties of radio emission also provides a unique diagnostic tool for 
recognizing gas flows and their distortions. 
In Sect.~\ref{s:observations} we describe the selection criteria of the constructed 
sample of interacting galaxies and present the used (VLA\footnote{The VLA of the NRAO is operated by Associated Universities, 
Inc., under cooperative agreement with the NSF}) radio data and the details of their 
reduction. The next section provides a brief description 
of each interacting system and its properties in radio total and polarized 
emission. We present estimations of equipartition magnetic field strength for all 
the galaxies and compare them with those for isolated spirals (Sect. \ref{s:magnetic}).

The statistical analysis of all the galaxies begins from investigating 
the evolution of magnetic fields as the interaction 
and merging process advance (\ref{s:evolution}). In Sect. \ref{s:sfr} we examine the relation 
of magnetic fields to the star formation rate (SFR). An analysis of observed morphological 
asymmetries in total and polarized radio emission is provided in Sect. \ref{s:asymmetries}.
The radio-far-infrared (FIR) diagram for interacting galaxies is presented in Sect. \ref{s:radiofir} 
and compared to the literature results. We also discuss the possible magnetization of IGM by magnetic fields of merger origin 
and the propagation effects of UHECRs in the vicinity of colliding galaxies (Sect \ref{s:impact}). 
We summarize our findings in Sect.~\ref{s:summary}.

\begin{table*}
\caption{Interacting systems and their members.}
\begin{center}
\begin{tabular}{lccccccc}
\hline\hline
Name                 & System name         &     Type           & Dist. & Inkl. & Pos. angle & \ion{H}{i} extent & Interaction \\
                     & or other name       &                    & [Mpc] & [deg.]&  [deg]     & [kpc]             & Stage       \\ 
\hline
\object{NGC\,876}$^S$         & NGC\,876/877        & SAc: sp            & 50.8  & 77.9  & 27.3       &  N/A              & $-1$        \\
\object{NGC\,877}$^S$         & NGC\,876/877        & SAB(rs)bc LIRG     & 50.8  & 35.6  & 138.0      &                   & $-1$        \\ 
\object{NGC\,4254}            & The Virgo Cluster   & SA(s)c,LINER,HII   & 17    & 32.0  & 60.0       & 40                & $-1$        \\
\object{NGC\,2207}$^S$        & NGC\,2207/IC\,2163  & SAB(rs)bc pec      & 35.0  & 58.2  & 115.9      &  56               & $-1$        \\
\object{IC\,2163}$^S$         & NGC\,2207/IC\,2163  & SB(rs)c pec        & 35.0  & 78.5  & 102.6      &                   & $-1$        \\  
\object{NGC\,5426}            & Arp\,271, NGC\,5426/5427 & SA(s)c pec    & 26.7  & 69.7  & 0.5        & 43                & $-1$        \\
\object{NGC\,5427}$^S$        & NGC\,5426/5427      & SA(s)c pec,Sy2,HII & 26.7  & 25.5  & 135.0      &                   & $-1$        \\
\object{NGC\,6907}$^S$        & NGC\,6907/6908      & SB(s)bc            & 44.5  & 37.5  & 66.7       & 78                & $0$        \\
\object{NGC\,6908}            & NGC\,6907/6908      & S                  & 44.5  &  N/A    & N/A      &                   & $0$        \\
\object{UGC\,12914}$^S$       & Taffy galaxy        & SAB(rs)c,Sbrst,Sy2 & 59.6  & 54.1  & 159.6      & 52                & $1$        \\
\object{UGC\,12915}           & Taffy galaxy        & Sdm                & 59.6  & 72.9  & 135.4      &                   & $1$        \\
\object{UGC\,813}             & Taffy2 galaxy       & Sb                 & 67    & 72.0  & 110.3      & 73                & $1$        \\
\object{UGC\,816}             & Taffy2 galaxy       & Sc                 & 67    & 62.0  & 170.0      &                   & $1$        \\
\object{NGC\,660}             &                     & SB(s)a pec,HII     & 12.3  & 78.8  & 11.9       & 47                & $1$        \\
\object{NGC\,4038}$^{STX}$    & Arp\,244,The Antennae & SB(s)m pec       & 26.8  & 51.9  & 133.2      & 133               & $1$        \\
\object{NGC\,4039}$^{STX}$    & The antennae        & SA(s)m,pec,LINER   & 26.8  & 71.2  & 132.0      &                   & $1$        \\
\object{NGC\,6621}$^T$        & Arp\,81             & Sb pec,LIRG        & 86.4  & 70.8  & 142.5      & 63                & $5$        \\
\object{NGC\,6622}$^T$        & Arp\,81             & Sa                 & 86.4  & 27.3  & 117.0      &                   & $5$        \\
\object{NGC\,520}$^{ST}$      & Arp\,157            & pec                & 30.2  & 77.4  & 130.0      & 103               & $7$        \\
\object{NGC\,3256}$^{TX}$     &                     & Sb(s) pec,LIRG     & 56    & 48.7  & 83.2       & 127               & $9$        \\
\object{Arp\,220}$^X$         &                     & S,Sy,ULIRG         & 76    & 57.0  & 96.5       & 133               & $10$        \\
\object{NGC\,7252}$^{TX}$     & Arp\,226,The Atoms of Peace  & (R)SA(r)  & 63    & 25.1  & 127.0      & 214               & $11$        \\
\object{Arp\,222}$^X$         &                     & SAB(s)a pec        & 23    & 43.4  & 66.5       & N/A               & $12$        \\
\object{NGC\,1700}$^X$        &                     & E4                 & 54    & 90.0  & 87.0       & N/A               & $13$        \\
\hline                                                                                                           
\end{tabular}
\end{center}
{\bf Notes.} $^{(T)}$ - a member of the Toomre sequence (\cite{toomre77}); $^{(X)}$ - a member of the 
X-ray sample of Brassington et al. (\cite{brassington07}); $^{(S)}$ - a member of our compiled sample 
of angularly-large interacting galaxies
\label{t:sample}
\end{table*}

\section{The sample and radio data reduction}
\label{s:observations}

\begin{table*}
\caption{Parameters of the reduced radio observations at 4.86\,GHz.}
%\begin{narrow}{-0.75in}{0in}
\begin{center}
\begin{tabular}{lcccccccc}
\hline
\hline
Name & Project ID & Amp. cal. & Phas. cal. & Integr. time & S$_{TP}$ & S$_{PI}$ & r.m.s. (TP) & r.m.s. (PI) \\
 & & & & [h] & [mJy] & [mJy] & [$\mu$Jy/beam] & [$\mu$Jy/beam] \\
\hline
NGC\,660 & AL296 & 3C138 & 0202$+$149 & 4.0 & 164.7$^{+1.1}_{-0.2}$ & 2.9$^{+1.0}_{-0.1}$ & 26.4 & 13.4 \\
NGC\,876/7 & AH314 & 3C48 & 0202$+$149 & 5.2 & 45.0$^{+0.5}_{-0.3}$ & 2.1$^{+1.2}_{-0.1}$ & 16.5 & 14.0 \\
NGC\,2207/IC\,2163 & AH370 & 3C286 & 0607$-$157 & 5.0 & 119.7$^{+0.9}_{-0.7}$ & 10.5$^{+0.9}_{-0.1}$ & 40.0 & 21.2 \\
NGC\,5426/7 & AL425 & 3C286 & 1406$-$076 & 3.9 & 44.9$^{+0.9}_{-0.3}$ & 3.8$^{+0.8}_{-0.1}$ & 23.3 & 14.9 \\
NGC\,6907/8 & AH314 & 3C286 & 2128$-$208 & 5.2 & 50.3$^{+0.3}_{-0.2}$ & 2.7$^{+0.5}_{-0.1}$ & 32.0 & 23.3 \\
TAFFY & AC264  & 3C48       & 0007+171 & 1.4 & 38.4$^{+0.4}_{-0.3}$ & 1.0$^{+0.7}_{-0.1}$ & 17.0 & 16.0 \\
TAFFY2 & AC592 & 3C48/3C138 & 0114+483 & 5.4 & 19.2$^{+1.1}_{-0.1}$ & 0.4$^{+0.6}_{-0.1}$ & 23.0 & 19.0 \\
\hline
\end{tabular}
\end{center}
%\end{narrow}
\label{t:reduced}
\end{table*}

\begin{table*}
\caption{Parameters of the original images from the NRAO VLA Archive$^{*}$}
%\begin{narrow}{-0.75in}{0in}
\begin{center}
\begin{tabular}{lcccccc}
\hline
\hline
Name & Project ID & Frequency & VLA config. & Integr. time & S$_{TP}$ & r.m.s. (TP) \\
 & & [GHz] & & [h] & [mJy] & [$\mu$Jy/beam] \\
\hline
NGC\,520 & AY102 & 1.43 & C & 1.0 & 172.8$^{+1.1}_{-5.6}$ & 150.0 \\
NGC\,1700 & NVSS & & D & & $<5.0$ & 350.0 \\
NGC\,3256 & AS412 & 4.86 & BC & 0.2 & 250.0$^{+11.3}_{-4.7}$ & 265.0 \\
NGC\,6621/2 & AU23 &  1.49 & BC & 0.1 & 28.0$^{+5.4}_{-0.6}$ & 225.0 \\
NGC\,7252 & AN090 & 4.86 & BC & 1.3 & 9.1$^{+0.9}_{-0.2}$ & 34.7 \\
Arp\,220 & AY102 & 1.43 & C & 0.6 & 334.3$^{+0.5}_{-2.2}$ & 175.0 \\
Arp\,222 & & 1.49$^{**}$ & D & & $2.9^{+0.7}_{-0.1}$ & $110.0$ \\
\hline
\end{tabular}
\end{center}
%\end{narrow}
{\bf Notes.} $^{(*)}$ from http://www.aoc.nrao.edu/\~{}vlbacald/; 
$^{(**)}$ from Condon (\cite{condon87})
\label{t:notreduced}
\end{table*}

To explore the evolution of magnetic field in interacting galaxies, we have compiled 
a sample of suitable objects. 
First, we chose objects of the best-known merging galaxy sequence of Toomre (\cite{toomre77}) for 
which appropriate radio data are available. We restricted our search to the archive of the VLA, 
frequencies 4.86 and 1.4 GHz, and
D, C, or BC configurations of the array to maximize the sensitivity for extended 
radio emission. There are 5 out of 11 merging systems with relevant radio data 
(see Table~\ref{t:sample}). Next, we selected six out of nine objects with available radio data from 
the X-ray merging sequence of Brassington et al. (\cite{brassington07})  (Table~\ref{t:sample}). 
Furthermore, we performed a more general search for angularly-large interacting 
objects with the usable radio data. We looked for galaxies to fulfil the following criteria:

\begin{itemize}

\item are RC3 galaxies (de Vaucouleurs et al. \cite{devaucouleurs91}) 
recognized as interacting objects according to Surace et al. (\cite{surace04});

\item are detected in the infrared (e.g. are included in the IRAS Bright Galaxies Sample: 
Sanders et al. \cite{sanders03});

\item at least one object in the interacting system has an angular size larger 
than $2\arcmin$ to study distribution of the radio emission, and are also smaller 
than $9\arcmin$ to be within the primary beam of the VLA antenna;

\item declination $>-40\degr$ ;

\item total flux of at least one galaxy in the interacting system is larger 
than 80\,mJy at 1.4\,GHz (from the NVSS, Condon et al. \cite{condon98}).

\end{itemize}

\begin{figure*}
\begin{minipage}[t]{9cm}
\begin{center}
\includegraphics[angle=270,width=9cm]{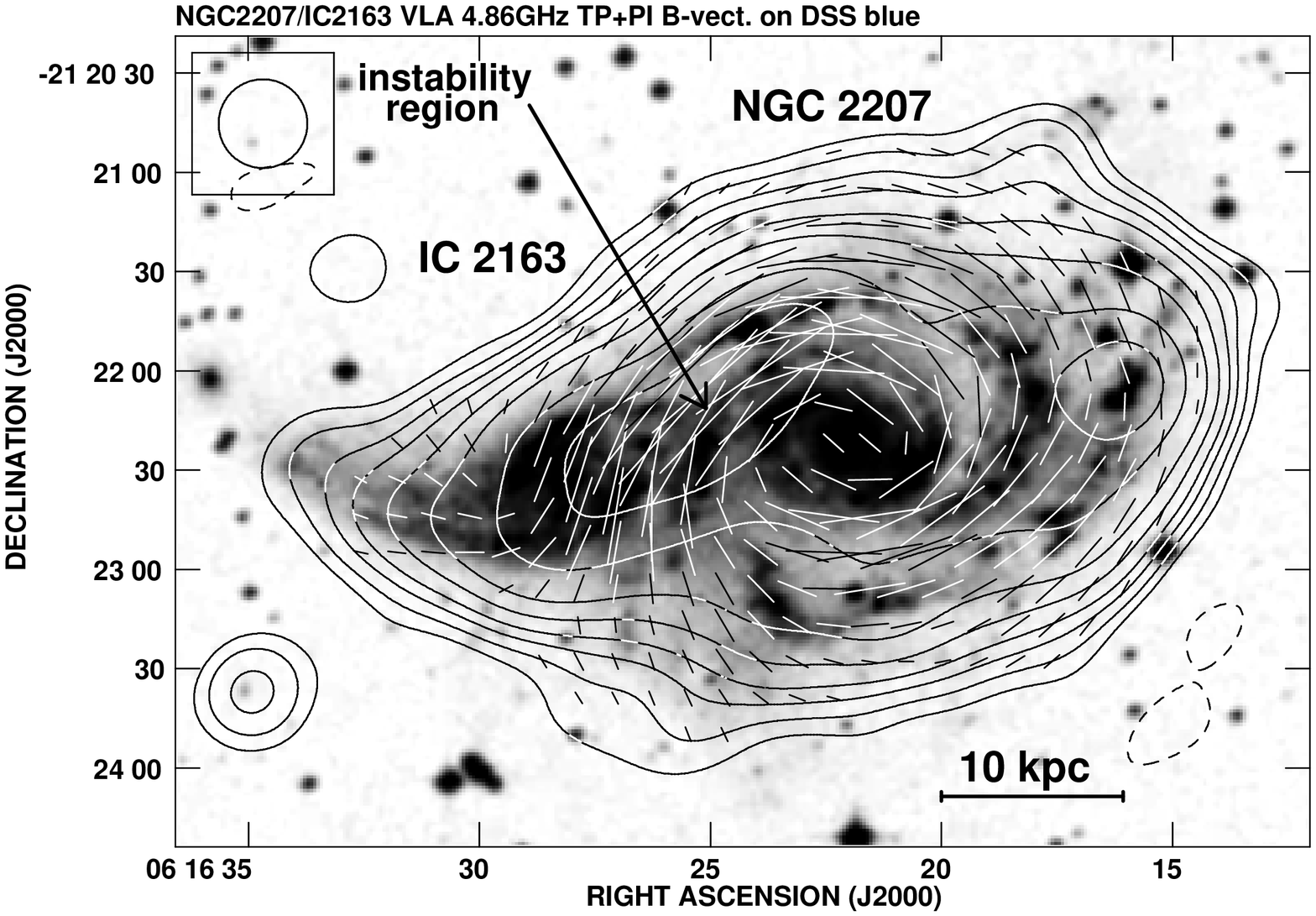}
\caption{The total-power contours and B-vectors of polarized intensity of NGC$\,$2207/IC$\,$2163 at 4.86$\,$GHz (natural weighting)
superimposed on the DSS blue image. The contour levels are (-3, 3, 5, 8, 12, 20, 35, 80, 150) $\times$ 40.0 $\mu$Jy/beam.
A vector of 10$"$ length corresponds to the polarized intensity of 166.7 $\mu$Jy/beam. The map resolution is 27$"$ $\times$ 27$"$ HPBW.}
\label{f:n2207tp}
\end{center}
\end{minipage}
\hfill
\begin{minipage}[t]{9cm}
\includegraphics[angle=270,width=9cm]{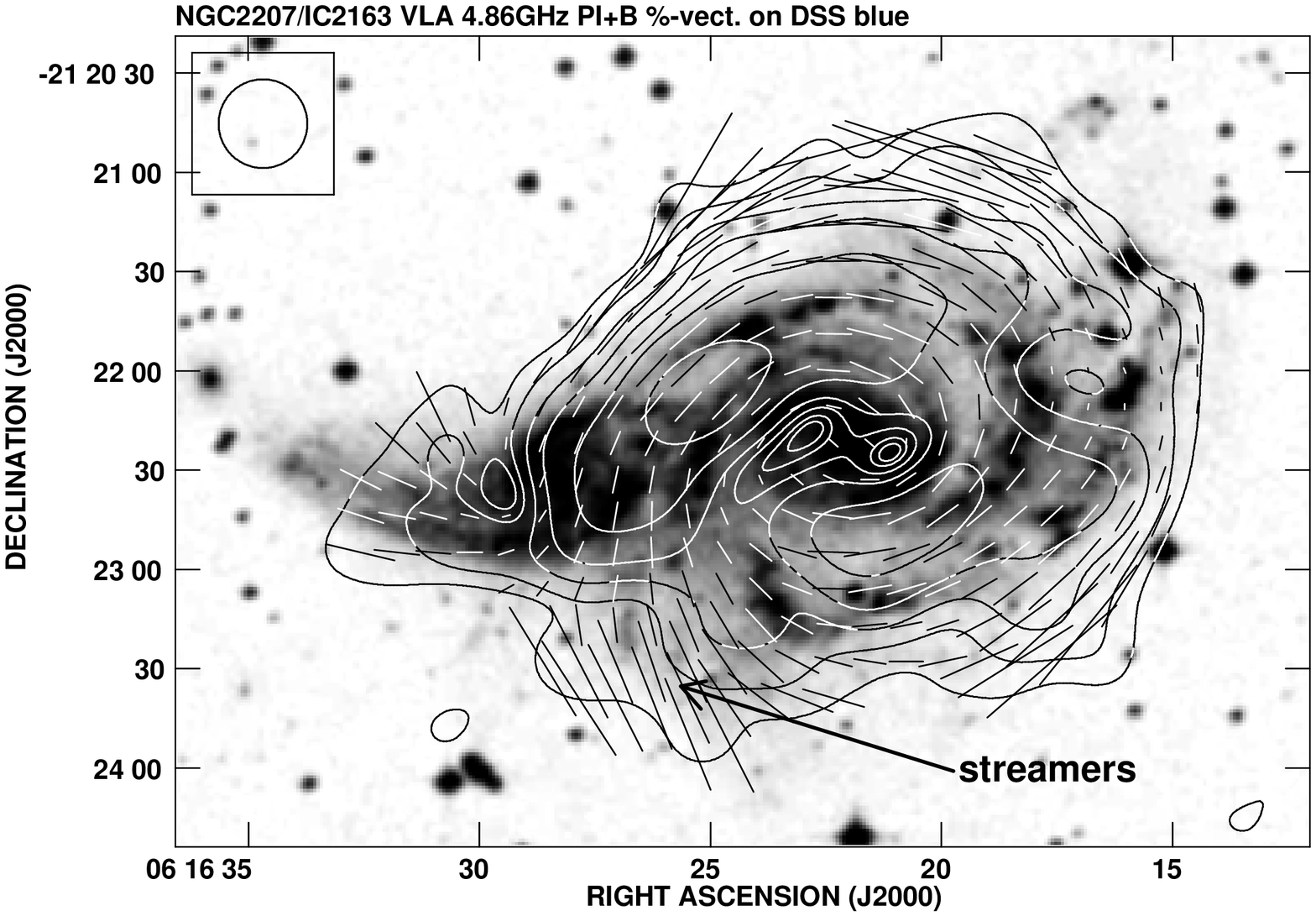}
\caption{The contours of polarized intensity and B-vectors of polarization degree of NGC$\,$2207/IC$\,$2163 at 4.86$\,$GHz
(natural weighting) superimposed on the DSS blue image. The contour levels are (3, 5, 8, 12, 20, 35) $\times$ 21.2 $\mu$Jy/beam. A
vector of 10$"$ length corresponds to the polarization degree of 10\%. The map resolution is 27$"$ $\times$ 27$"$ HPBW.}
\label{f:n2207pi}
\end{minipage}
\end{figure*}

We obtained 24 objects satisfying those requirements. For ten objects, which are members 
of seven interacting galaxy pairs, we found high-quality VLA archive radio polarimetric data at 4.86\,GHz 
(see Table\,\ref{t:sample}). We enriched this sample by a few 
important types of interacting objects: the 
polar-ring galaxy NGC\,660, the weakly interacting spiral NGC\,4254 in the 
Virgo Cluster (see Chy\.zy et al. \cite{chyzy07b}, Chy\.zy \cite{chyzy08}), 
and UGC\,813/UGC\,816 (see Sect.~\ref{s:taffy2}). In total, in our final sample 
we assembled 24 pre- and post-merger galaxies from 16 recognized interacting systems.
Although they by no means constitute a complete sample of colliding galaxies, 
they do represent the different stages of tidal interaction and merging process, which is suitable for 
our study. 

We reduced the original VLA radio data for the seven interacting systems listed 
in Table~\ref{t:reduced}. Some of them were unpublished to date, some were 
published but without any analysis of polarization data, and for some objects we 
obtained the final images of better quality than those known from the 
literature (see the next section). Using the standard AIPS package and procedures, 
the instrumental polarization was corrected with phase calibrators, which were
also applied in gain and phase calibration (Table~\ref{t:reduced}).
The flux density scale was established by the amplitude calibration sources 
and the data were self-calibrated in phase, except for NGC\,5426/7. NGC\,6907/8 data were 
also self-calibrated in amplitude. 
The combined maps in I, Q, and U Stokes parameters at both
frequencies were then used in constructing the maps of polarized intensity,
polarization position angle, and position angle of observed magnetic vectors
(observed $E$--vectors rotated by $90\degr$ -- the so-called $\vec{B}$-vectors).
The details of observations, obtained r.m.s. sensitivities of the radio maps at 4.86\,GHz, and 
total and polarized radio fluxes are presented in Table~\ref{t:reduced}.

Also at our disposal were reduced high-quality VLA radio polarization data 
at 4.86\,GHz for the Antennae galaxies and for the Virgo Cluster member NGC\,4254 
from our previous publications (see for details Chy\.zy \& Beck 
\cite{chyzy04}; Chy\.zy et al. \cite{chyzy07b}; Chy\.zy \cite{chyzy08})).
For the seven other mergers we used the original images from the VLA archive. The 
details of those data and r.m.s. sensitivities in total radio emission 
are presented in Table~\ref{t:notreduced}. The short integration time 
prevented us from calibrating the polarized emission for these seven objects.

\section{Results}
\label{s:results}

\begin{figure*}
\begin{minipage}[t]{9cm}
\begin{center}
\includegraphics[angle=0,width=8cm]{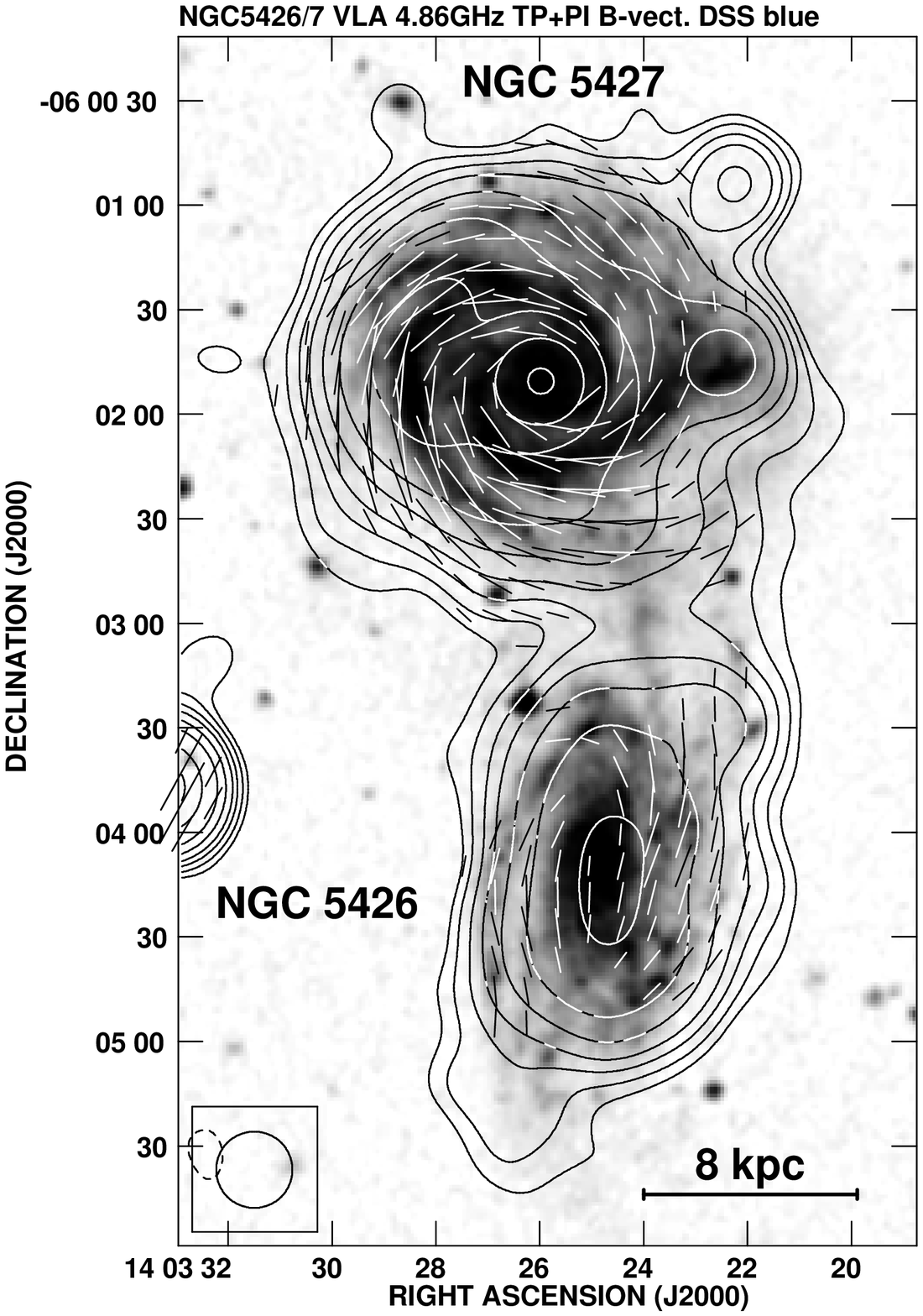}
\caption{The total power contours and B-vectors of polarized intensity of NGC$\,$5426/NGC$\,$5427 at 4.86$\,$GHz (natural weighting)
superimposed on the DSS blue image. The contour levels are (-3, 3, 5, 8, 12, 20, 35, 80, 150, 250) $\times$ 23.3 $\mu$Jy/beam.
A vector of 10$"$ length corresponds to the polarized intensity of 83.3 $\mu$Jy/beam. The map resolution is 22$"$ $\times$ 22$"$ HPBW.}
\label{f:n5426tp}
\end{center}
\end{minipage}
\hfill
\begin{minipage}[t]{9cm}
\includegraphics[angle=0,width=8cm]{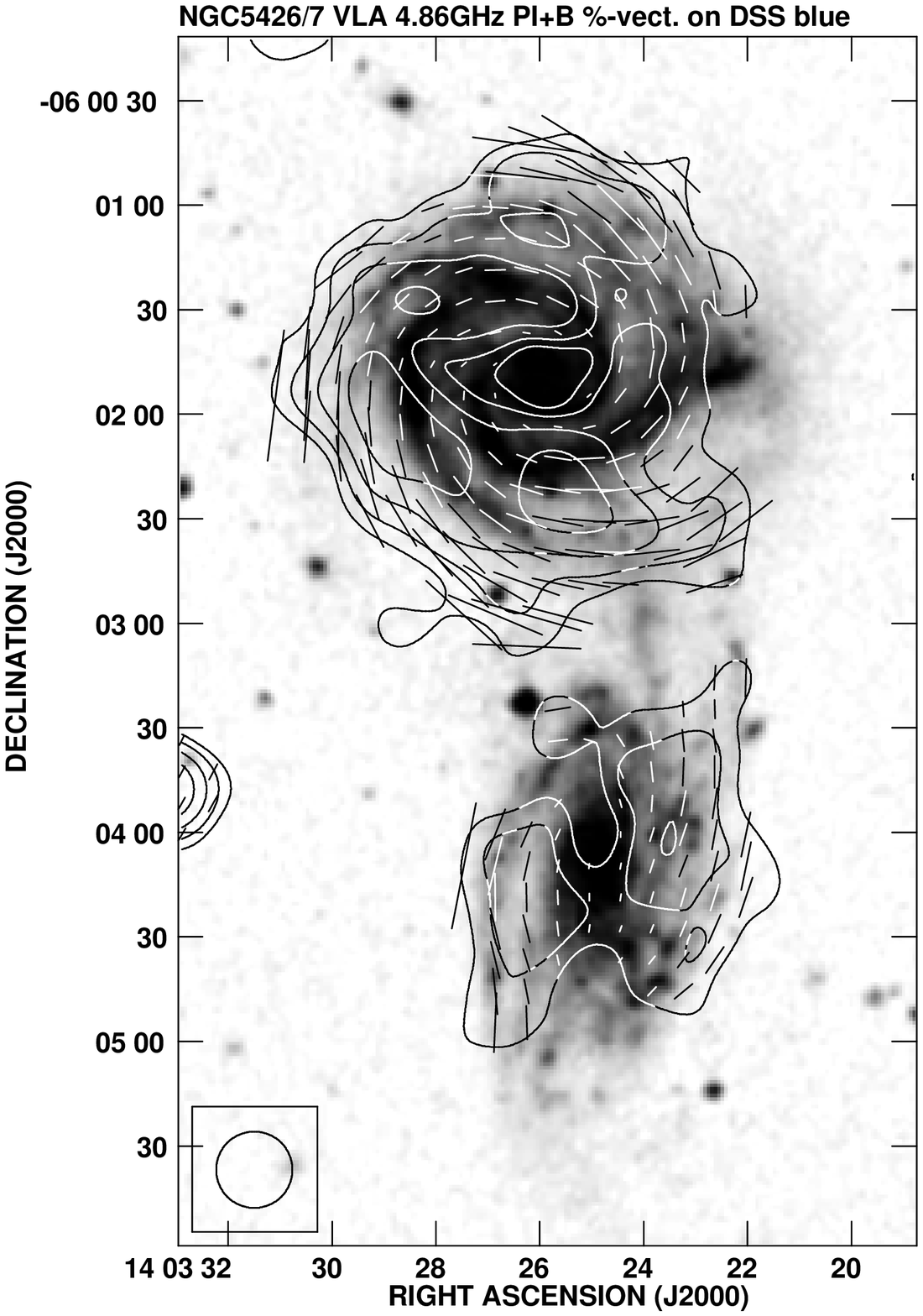}
\caption{The contours of polarized intensity and B-vectors of polarization degree of NGC$\,$5426/NGC$\,$5427 at 4.86$\,$GHz
(natural weighting) superimposed on the DSS blue image. The contour levels are (3, 5, 8, 12) $\times$ 14.9 $\mu$Jy/beam. A
vector of 10$"$ length corresponds to the polarization degree of 17\%. The map resolution is 22$"$ $\times$ 22$"$ HPBW.}
\label{f:n5426pi}
\end{minipage}
\end{figure*}

In this section we describe radio total and polarized emission 
of our interacting galaxies and compare 
them with morphological properties in other wavelengths. A detailed presentation 
of resolved systems is followed by concise descriptions 
of objects without well-resolved radio structures. Next, we estimate magnetic field strengths for 
all the objects investigated.

\subsection{NGC\,2207/IC\,2163}
\label{s:2207}

In this pair of spiral galaxies, NGC\,2207 appears in the foreground, obscuring the other one,  IC\,2163. 
According to the numerical simulations, $4\times10^7$\,yr 
ago IC\,2163 passed behind NGC\,2207 from west to east at a distance of about 20\,kpc 
(Elmegreen et al. \cite{elmegreen95b}). Seventeen young stellar clusters were 
discovered in both galaxies (Elmegreen et al. \cite{elmegreen01}), but the 
interaction has not yet led to a prominent starburst.

The radio emission at 4.86\,GHz (Fig. \ref{f:n2207tp}) covers the optical extent of the 
system but displays a hole in the central part of NGC\,2207 like in the map at 1.4\,GHz (Condon \cite{condon83}). 
The emission traces the optical western tidal tail of IC\,2163 and peaks at the eastern edge
of NGC\,2207, where a gravitational instability region is suspected, producing 
a maximum in the infrared emission (Elmegreen et al. \cite{elmegreen06}).
In this region, the total and polarized emission also have maxima, but not that degree of 
polarization, so there is no enhanced production of ordered magnetic field here, and 
the found radio properties of this region could result from a supply of relativistic 
electrons coming from the western tidal tail of the IC$\,$2163. This idea 
is supported by accretion of gas from the IC$\,$2163 onto the NGC$\,$2207, as suggested by 
Elmegreen et al. (\cite{elmegreen95a}). The \ion{H}{i} velocity dispersion 
is quite high in a large area of the disk of NGC\,2207 ($40-50\,$km\,s$^{-1}$).
However, we do not see any correspondence of regions with enhanced velocity dispersion 
(cf. Elmegreen et al. \cite{elmegreen95a}) with observed distribution of total or 
polarized radio emission. Therefore, the tidally generated turbulence and/or stretching 
motions along the line of sight required to explain the observed pattern of velocity 
dispersion cannot significantly influence the magnetic field strength.

The regular magnetic field around the central part of NGC\,2207 
have an almost exclusively azimuthal direction (Fig. \ref{f:n2207pi}). Also HI gas forms a similar 
ring-like structure (Elmegreen et al. \cite{elmegreen95a}). However, in the southern 
part of the disk, the $\vec{B}$-vectors have a larger pitch angle ($~45\degr$) and 
go out of the disk. This may indicate an outflow of magnetized plasma along a tidal tail and conversion 
of random magnetic field to the regular component as in the eastern tail of the 
Antennae galaxies (Chy\.zy \& Beck \cite{chyzy04}). 
With rising distance from the disk, the degree of polarization rises 
to 40\%, and the magnetic field becomes more regular. In this region, the optical and infrared emission 
form long and narrow structures called ``streamers'' and have the same pitch angle 
as $\vec{B}$-vectors (Fig. \ref{f:n2207pi}). In this part of NGC\,2207, a large 
bubble of \ion{H}{i} gas extends up to 30\,kpc out of the disk  to the south (Elmegreen 
et al. \cite{elmegreen95a}). 

A similar behaviour of the magnetic field is also observed in the eastern part of IC\,2163.
The degree of polarization becomes larger with distance from the disk centre
(reaching up to 30\%), which can indicate stretching of field lines along the tidal tail 
(Fig. \ref{f:n2207pi}).

\subsection{NGC\,5426/7}

Both galaxies in this binary system are spirals of the same Hubble class 
Sc, with similar sizes ($\approx 20$\,kpc), and masses ($~3\times10^{10}$\,M$\sun$;
Fuentes-Carrera et al. \cite{fuentes04}).
The evidence for weak gravitational interaction 
are the thin straight strands of the outer arms of NGC\,5426 
that overlap the southern spiral arms of NGC\,5427 and the straight 
arm segment in the SW part of the spiral structure of NGC\,5427 (Fig. \ref{f:n5426tp}).
The velocity field of ionized gas in NGC\,5427 is also slightly distorted 
(Fuentes-Carrera et al. \cite{fuentes04}). 
Blackman (\cite{blackman82}) found that 38\% of the total optical flux 
of the system comes from two adjacent halves of both the disks. This 
suggests obscuration by dust in NGC\,5426 that situates this galaxy 
in front of NGC\,5427 (Fuentes-Carrera et al. \cite{fuentes04}).
All these features can be accounted for by an early phase of interaction that suggests a closer passage of galaxies in $5\times 10^7$\,yrs 
(Fuentes-Carrera et al. \cite{fuentes04}).

This scenario of weak tidal interaction fully agrees with the 
radio polarimetric data at 4.86\,GHz. The total radio intensity closely follows the optical morphology 
(Fig. \ref{f:n5426tp}). Both galaxies have coherent magnetic spiral patterns that closely follow 
the optical spiral arms. The absence of polarized emission in between the galaxies 
(Fig. \ref{f:n5426pi}) is probably due to beam depolarization of two structures of perpendicular 
$\vec{B}$-vectors that overlap there.

In the most disturbed SW part of NGC\,5427 there is no polarized emission, which makes 
this galaxy slightly asymmetric in the polarized intensity (Fig. \ref{f:n5426pi}).
In the SE part of NGC\,5427 the polarized emission occurs outside
the density wave. To date, such outer magnetic arms have only been observed 
in the weakly interacting Virgo cluster spiral NGC\,4254 (Chy\.zy et al. \cite{chyzy07b}). 

\begin{figure*}
\begin{minipage}[t]{9cm}
\begin{center}
\includegraphics[angle=0,width=8.5cm]{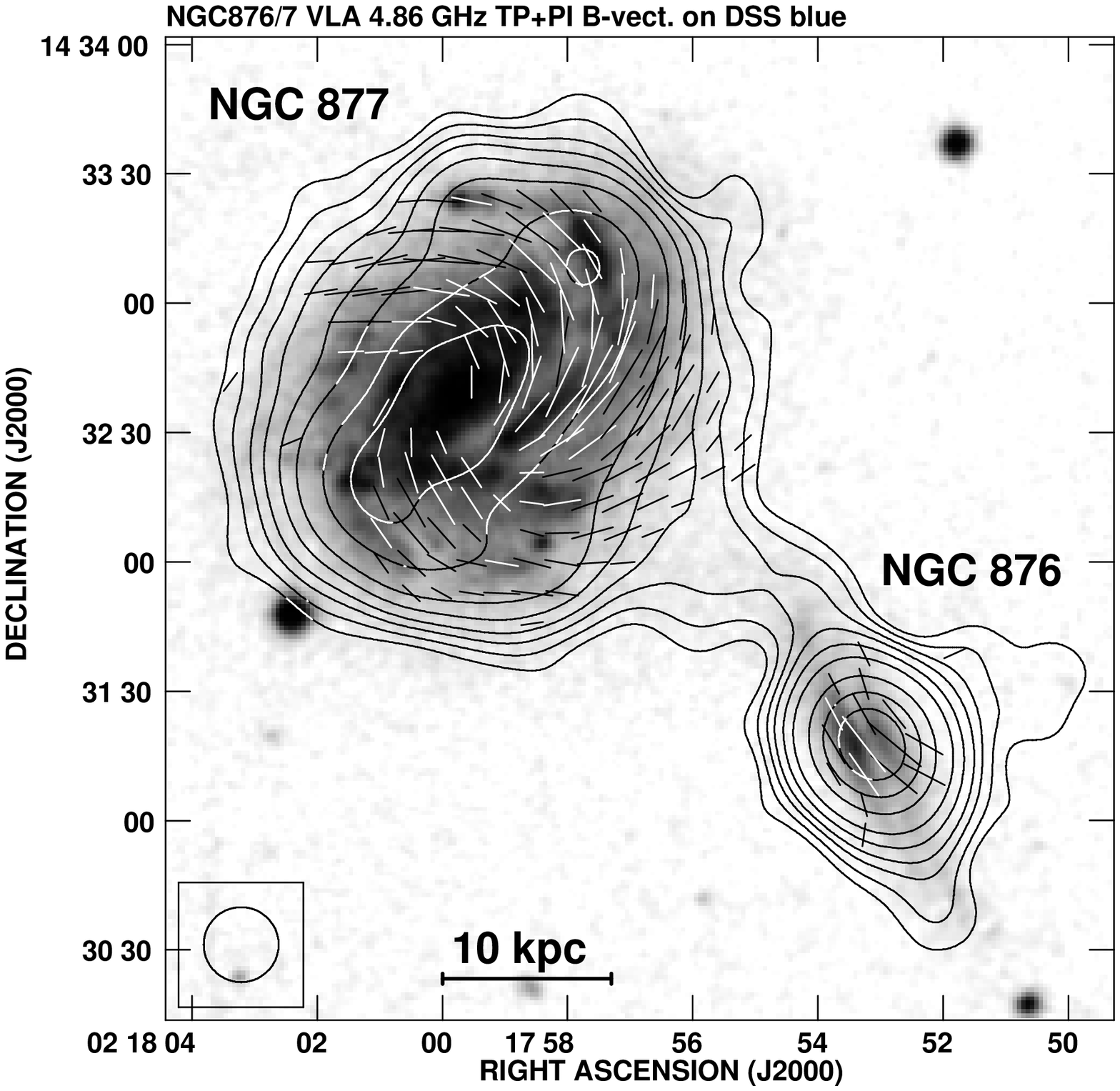}
\caption{The total power contours and B-vectors of polarized intensity of NGC$\,$876/NGC$\,$877 at 4.86$\,$GHz (natural weighting)
superimposed on the DSS blue image. The contour levels are (-3, 3, 5, 8, 12, 20, 35, 80, 150, 250) $\times$ 16.5 $\mu$Jy/beam.
A vector of 10$"$ length corresponds to the polarized intensity of 83.3 $\mu$Jy/beam. The map resolution is 18$"$ $\times$ 18$"$ HPBW.}
\label{f:n877tp}
\end{center}
\end{minipage}
\hfill
\begin{minipage}[t]{9cm}
\centering
\includegraphics[angle=0,width=8.5cm]{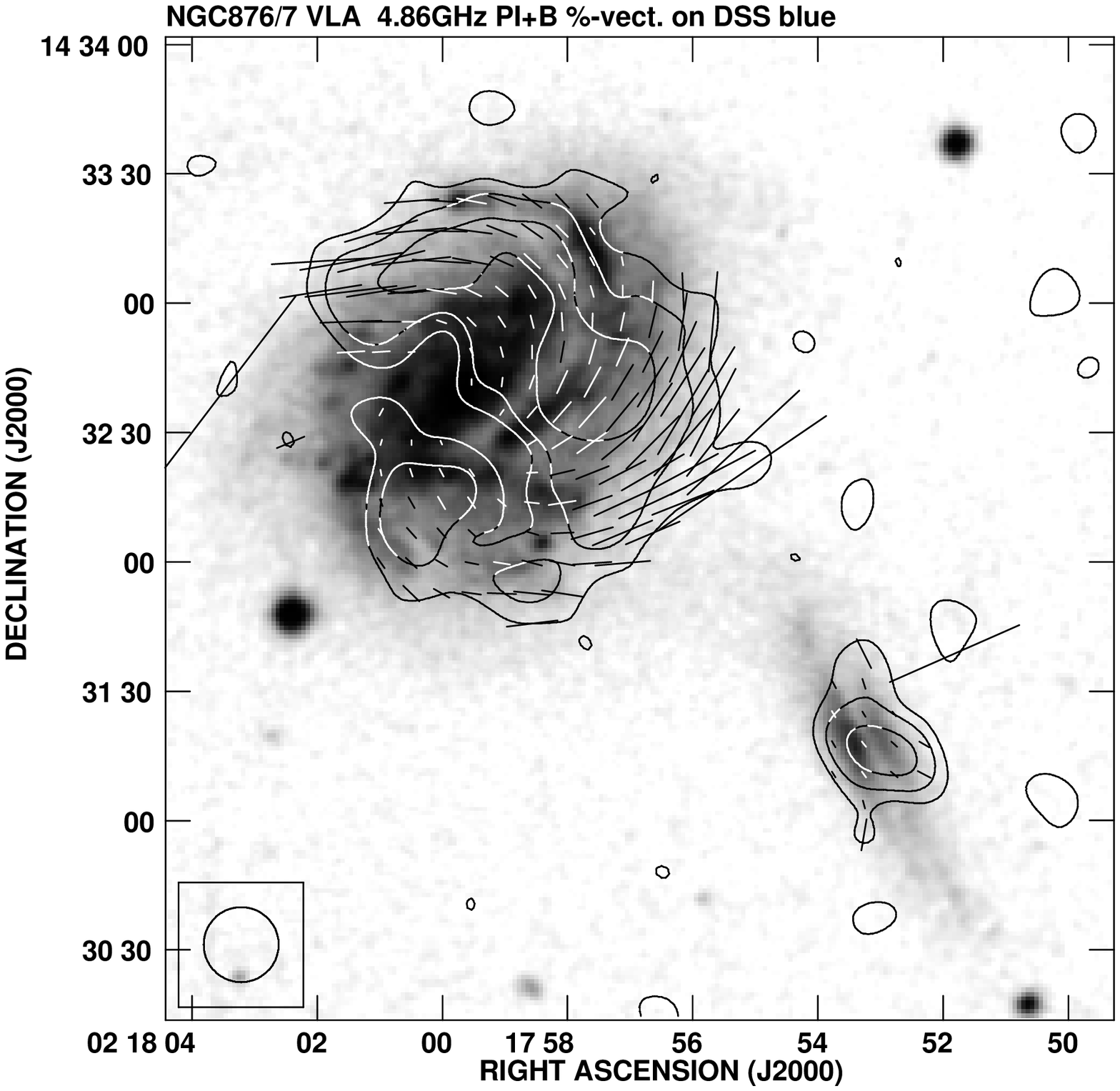}
\caption{The contours of polarized intensity and B-vectors of polarization degree of NGC$\,$876/NGC$\,$877 at 4.86$\,$GHz
(natural weighting) superimposed on the DSS blue image. The contour levels are (3, 5, 8) $\times$ 14.0 $\mu$Jy/beam. A
vector of 10$"$ length corresponds to the polarization degree of 14\%. The map resolution is 18$"$ $\times$ 18$"$ HPBW.}
\label{f:n877pi}
\end{minipage}
\end{figure*}

\subsection{NGC\,876/7}

The larger component of this pair of galaxies -- NGC\,877 -- is a late-type spiral 
with prominent and slightly disturbed optical spiral arms (Fig. \ref{f:n877tp}). 
Bright H$\alpha$ emitting regions are distributed along the arms but are not 
condensed in the central part of the disk (Hattori et al. \cite{hattori04}). The 
companion is a small, highly inclined galaxy ($78\degr$) at a projected separation of about 50\,kpc.

The total radio emission at 4.86\,GHz of both the galaxies corresponds well to optical and 
H$\alpha$ disks and even reveals a bridge of low-surface brightness emission 
(at $~180\,\mu$Jy level) between the components (Fig. \ref{f:n877tp}). 
The strong and elongated radio structure in the main body of NGC\,877 is 
a galactic bar, from which regular magnetic fields stretch out to the N and S.
The magnetic pattern is very coherent one and resembles those in grand-design spiral galaxies. 
The presence of tidally excited spiral arms in a system with a small companion is 
predicted by numerical simulations (Toomre \cite{toomre81}). This would explain 
the excess in infrared emission of NGC\,877, which led to classifying this object 
as LIRG (Sanders et al. \cite{sanders03}).
The smaller companion (NGC\,876) has more concentrated radio emission, which may result from
high inclination ($78\degr$) of this spiral. Surprisingly enough, it shows clear
hints of the X-shape structure of $\vec{B}$-vectors, often observed in the edge-on field galaxies 
(Fig. \ref{f:n877tp}). 

The distribution of the polarized emission of NGC\,877 is asymmetrical (Fig. \ref{f:n877pi}).
The major part of this emission comes from the NW part of the disk, where the spiral arm 
is also better visible in the optical images. The optical morphology in the SE part of 
the disk is only slightly distorted but corresponds to a distinct magnetic spiral arm. 
A blob of polarized emission in the SE part of the disk coincides with an appendage in 
the near-infrared 
emission, which also accounts for asymmetrical appearance of this object at this band (Moriondo et 
al. \cite{moriondo99}). The companion does not appear clearly asymmetrical in the polarized
emission, indicating a weaker effect from interaction onto this object. The weak distortions
of magnetic fields also indicate an early type of interaction of this system.

\subsection{NGC\,6907/8}

\begin{figure*}
\begin{minipage}[t]{9cm}
\begin{center}
\includegraphics[angle=0,width=7cm]{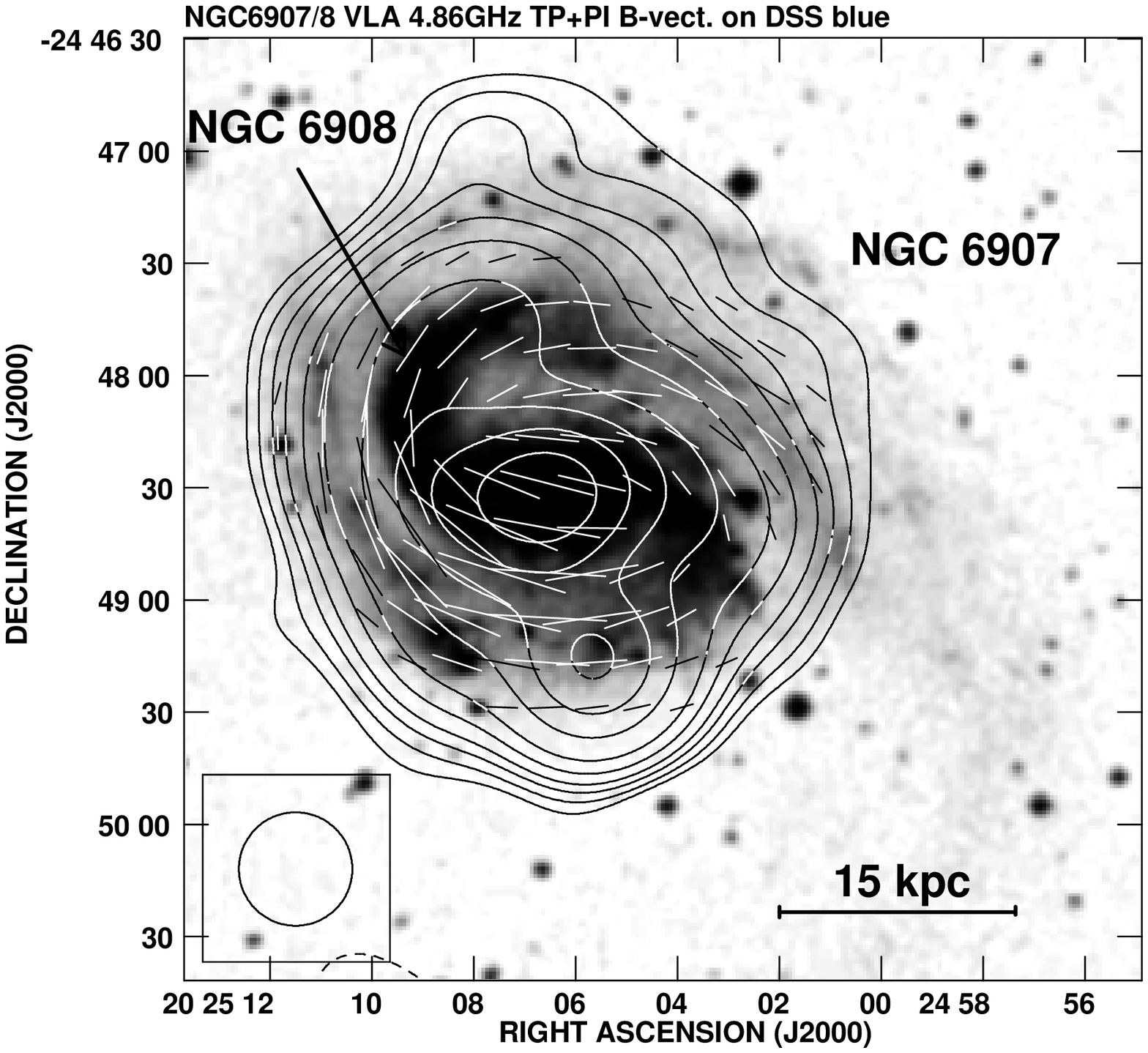}
\caption{The total power contours and B-vectors of polarized intensity of NGC$\,$6907/NGC$\,$6908 at 4.86$\,$GHz (natural weighting)
superimposed on the DSS blue image. The contour levels are (-3, 3, 5, 8, 12, 20, 35, 80, 150, 250, 400) $\times$ 32.0 $\mu$Jy/beam.
A vector of 10$"$ length corresponds to the polarized intensity of 125.0 $\mu$Jy/beam. The map resolution is 30$"$ $\times$ 30$"$ HPBW.}
\label{f:n6907tp}
\end{center}
\end{minipage}
\hfill
\begin{minipage}[t]{9cm}
\centering \includegraphics[angle=0,width=7cm]{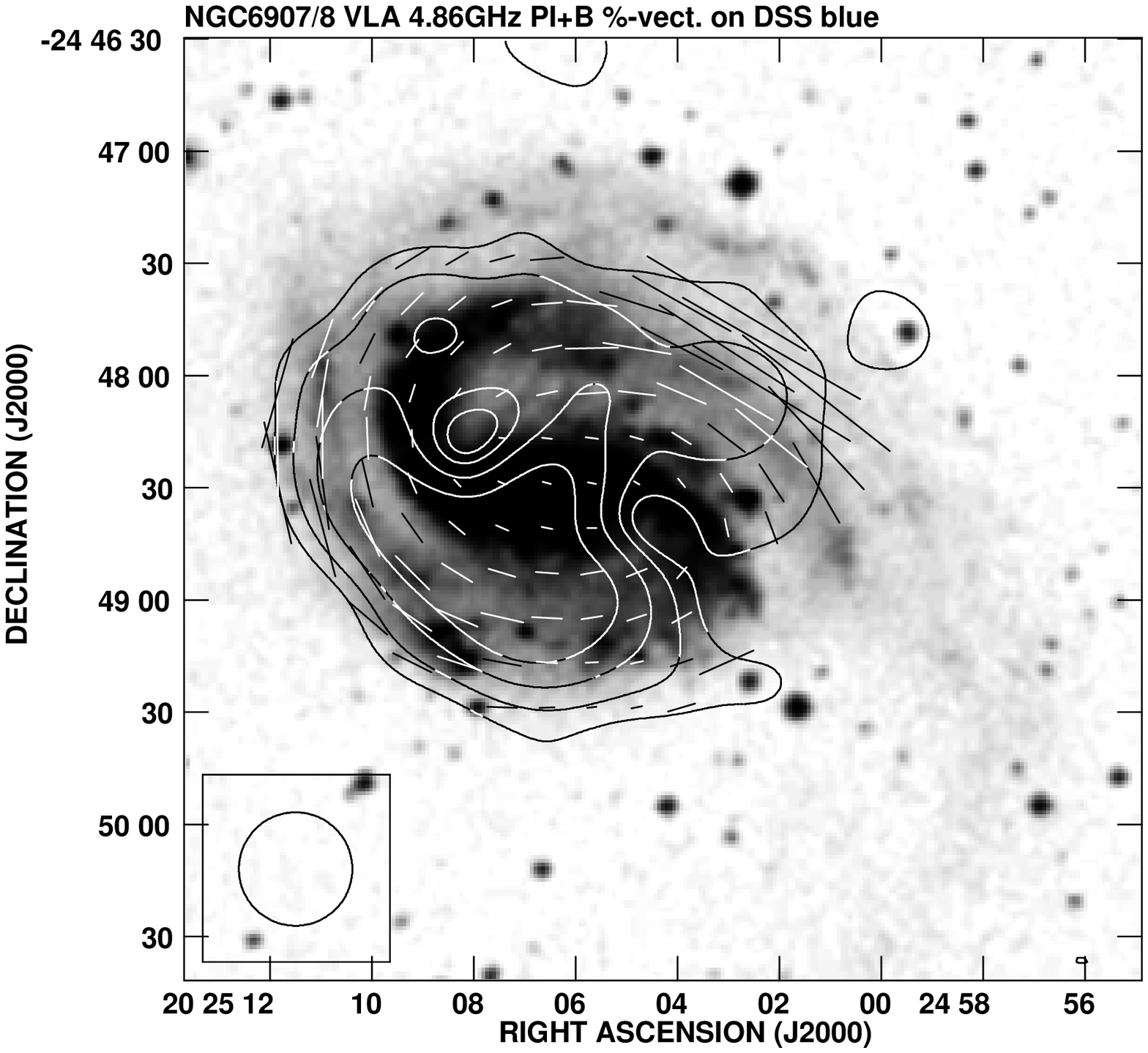}
\caption{The contours of polarized intensity and B-vectors of polarization degree of NGC$\,$6907/NGC$\,$6908 at 4.86$\,$GHz
(natural weighting) superimposed on the DSS blue image. The contour levels are (3, 5, 8, 12) $\times$ 23.3 $\mu$Jy/beam. A
vector of 10$"$ length corresponds to the polarization degree of 8\%. The map resolution is 30$"$ $\times$ 30$"$ HPBW.}
\label{f:n6907pi}
\end{minipage}
\end{figure*}

NGC\,6908 is a small lenticular galaxy superimposed on the NE arm of 
the dominant galaxy NGC\,6907. It is most clearly seen in near-IR images (Madore 
et al. \cite{madore07}). The spiral structure of NGC\,6907 is distorted: its 
eastern arm is much longer and wraps around the entire galaxy (Fig. \ref{f:n6907tp}).
The bulge of NGC\,6907 has an elliptical shape, and is shifted off the base of the arms. 
The analysis of distortions in the HI velocity field suggests that the first collision 
between galaxies occurred $3.4\pm 0.6\times 10^7$\,yr ago, leaving behind a triangular HI gas stream 
to the NE from the current position of NGC\,6908 (Scarano et al. \cite{scarano08}). 
Sanders et al. (\cite{sanders03}) classifies NGC\,6907 as LIRG.

In our maps at 4.86\,GHz, the total radio emission is detected only from the larger galaxy and 
concentrates in its central part (Fig. \ref{f:n6907tp}). Two protrusions to N and S are likely 
some background objects. The polarized emission (Fig. \ref{f:n6907pi}) is more asymmetric 
and dominates outside the galactic bar in the SE part of the disk. The polarized data reveal 
a coherent spiral structure of magnetic $\vec{B}$-vectors of small pitch angle (0\degr-30\degr), 
even in its eastern part, where the optical density waves are weak. In the eastern part of 
the disk the well-ordered magnetic fields seem to be pulled out and directed along the tidally 
stretched spiral arm.  

\subsection{The Taffy}
\label{s:taffy}

\begin{figure*}
\begin{minipage}[t]{9cm}
\begin{center}
\includegraphics[angle=0,width=8.0cm]{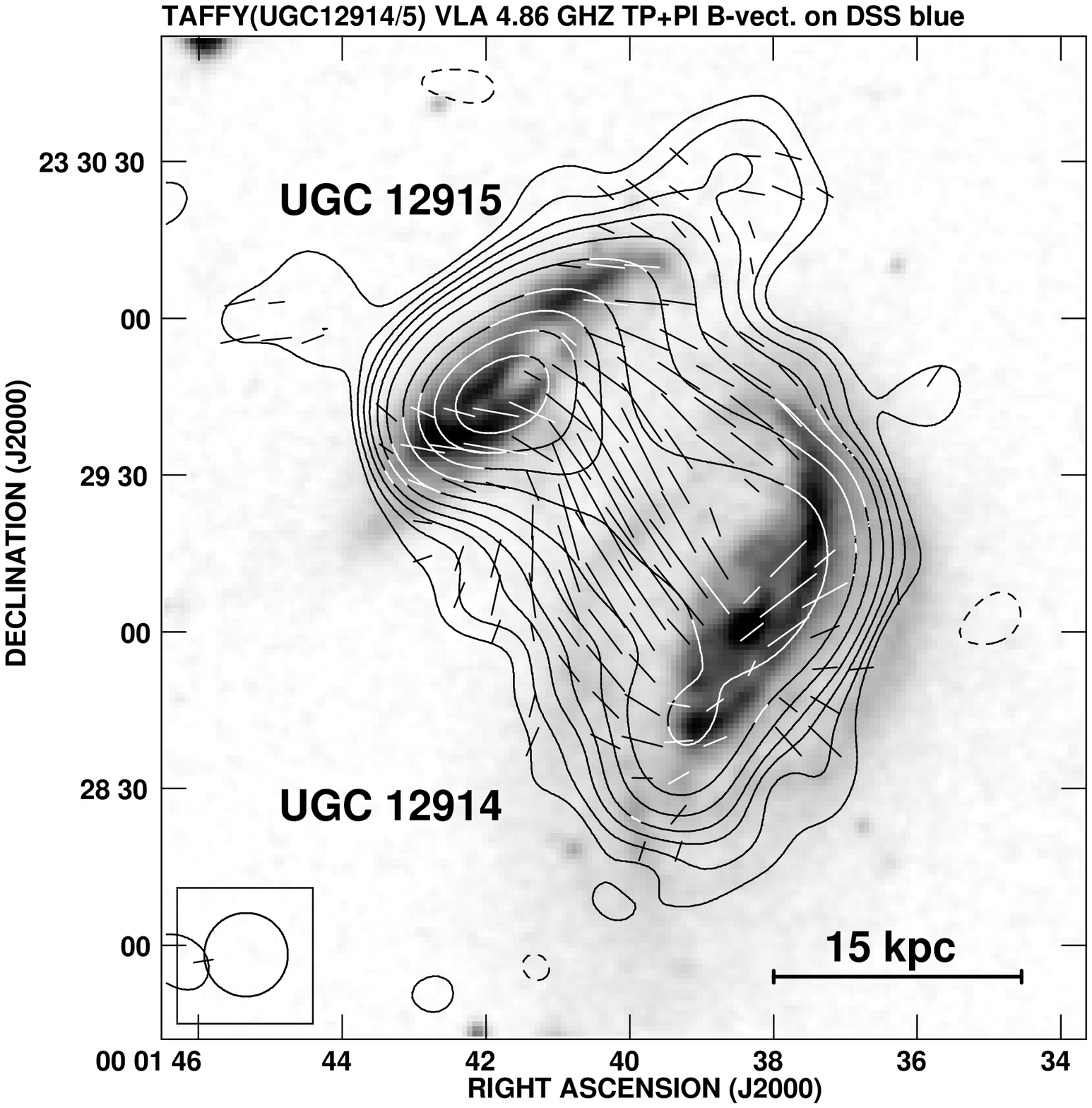}
\caption{The total power contours and B-vectors of polarized intensity of the Taffy (UGC$\,$12914/UGC$\,$12915) at 4.86$\,$GHz 
superimposed on the DSS blue image. The contour levels are (-3, 3, 5, 8, 12, 20, 35, 80, 150, 250, 400) $\times$
17.0 $\mu$Jy/beam. A vector of 10$"$ length corresponds to the polarized intensity of 71.4 $\mu$Jy/beam. The map resolution 
is 16$"$ $\times$ 16$"$ HPBW.}
\label{f:taffytp}
\end{center}
\end{minipage}
\hfill
\begin{minipage}[t]{9cm}
\begin{center}
\includegraphics[angle=0,width=9.0cm]{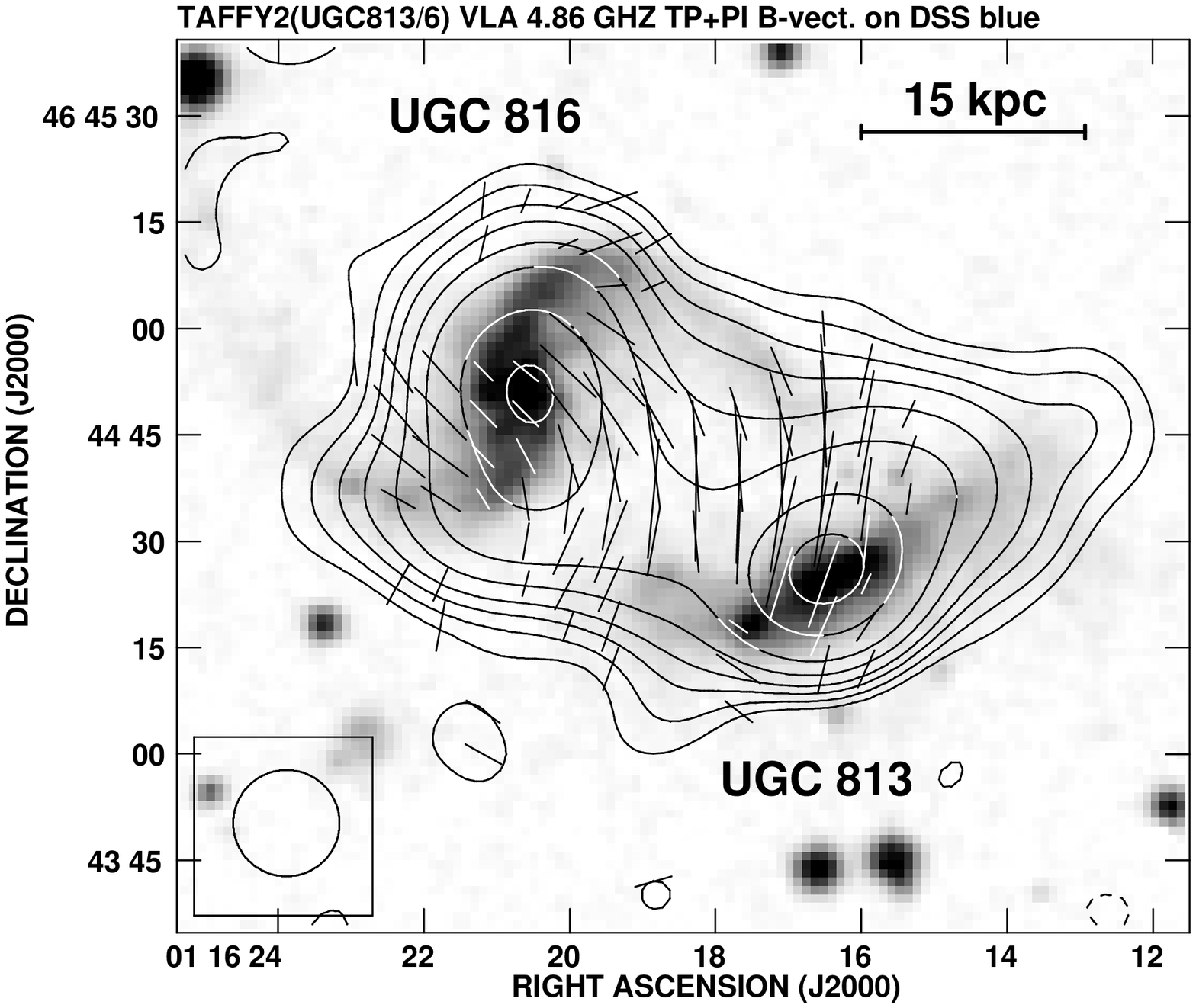}
\caption{The total power contours and B-vectors of polarized intensity of the Taffy2 (UGC$\,$813/UGC$\,$816) at 4.86$\,$GHz 
superimposed on the DSS blue image. The contour levels are (-3, 3, 5, 8, 12, 20, 35, 80, 150) $\times$
23.0 $\mu$Jy/beam. A vector of 10$"$ length corresponds to the polarized intensity of 69.4 $\mu$Jy/beam. The map resolution 
is 15$"$ $\times$ 15$"$ HPBW.}
\label{f:taffy2tp}
\end{center}
\end{minipage}
\end{figure*}

The unusual morphology of this pair of galaxies (UGC\,12914 and 
UGC\,12915), known as the Taffy system, comes from a recent 
($\approx 10^7$\,yr), nearly head-on encounter (Condon et al.\cite{condon93}).
Almost half of their radio emission at 1.49\,GHz have been detected 
just from the bridge between the galaxies, which may be due to fresh relativistic CR 
electrons radiating in magnetic fields pulled out of the disks. In this system 
the \ion{H}{i} gas is mostly distributed along the bridge (Condon et al. \cite{condon93}), where a large 
amount of molecular gas and dust were also detected (Gao et al. \cite{gao03}; 
Zhu et al. \cite{zhu07}).

Our re-reduced radio polarization data at 4.86\,GHz (Fig. \ref{f:taffytp}) 
show more details of the magnetic field structure than are presented in 
Condon et al. (\cite{condon93}). We have confirmed that the general orientation 
of magnetic field is along the line joining the galaxies, which is 
also a possible direction for passing the intruder galaxy 
UGC\,12914. In addition, we found that in UGC\,12915 the polarized emission is clearly 
visible on the interacting side of the disk, whereas the other side is 
almost completely devoid of polarized signal. Either the magnetic 
field was enhanced asymmetrically by the interaction or the magnetized ISM 
on the NE side of the disk was stripped away by ram pressure during the passage 
through UGC\,12914. The UGC\,12914 shows some patches of polarized 
signal on the disk side opposite the bridge, which are associated with ordered magnetic 
fields with large pitch angles of up to $90\degr$ (Fig. \ref{f:taffytp}). 

\begin{figure*}
\begin{minipage}[c]{9cm}
\centering
\includegraphics[angle=0,width=7cm]{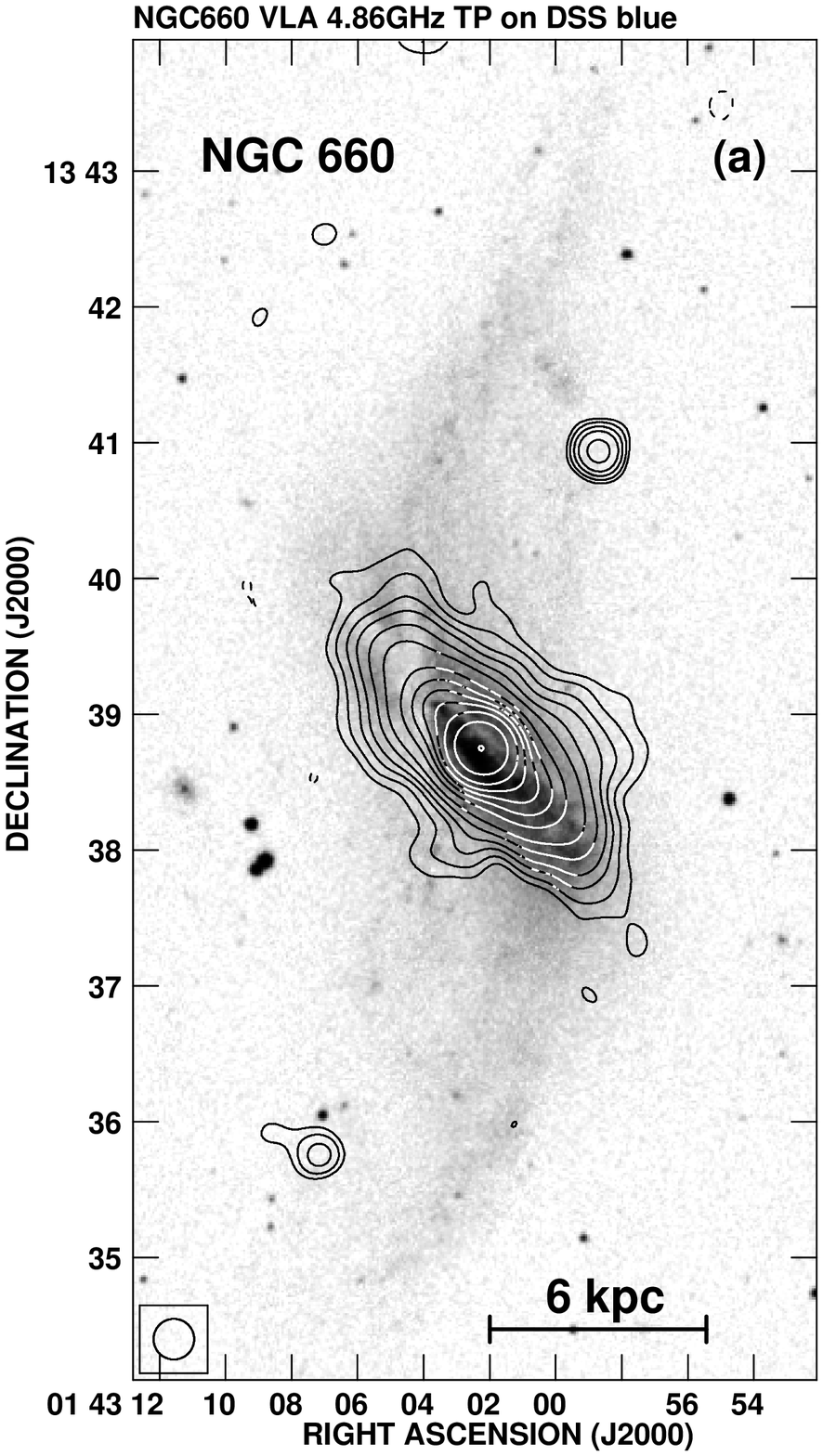}
\end{minipage}%
\begin{minipage}[c]{9cm}
\centering
\includegraphics[angle=0,width=5.5cm]{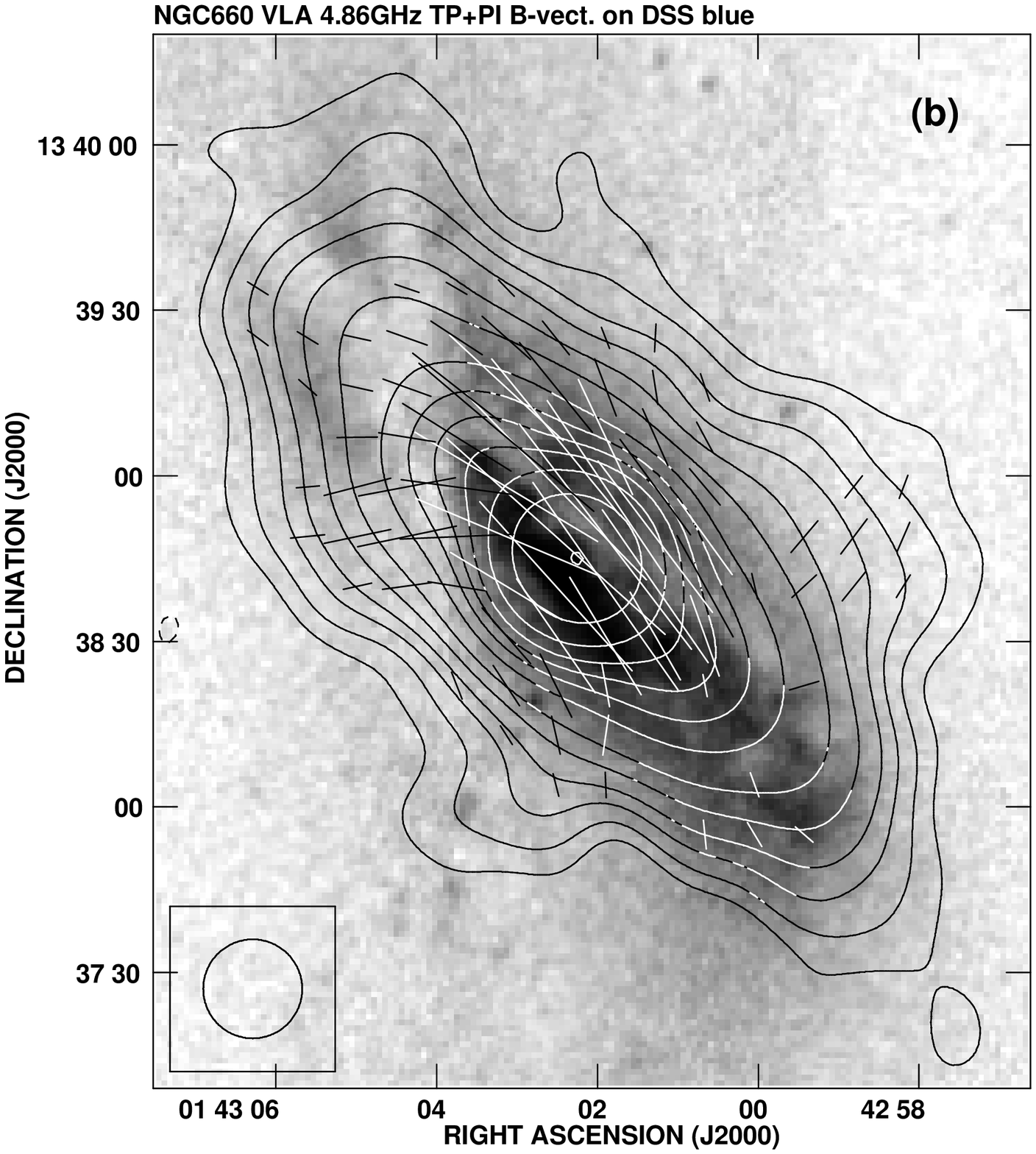}
\includegraphics[angle=0,width=5.5cm]{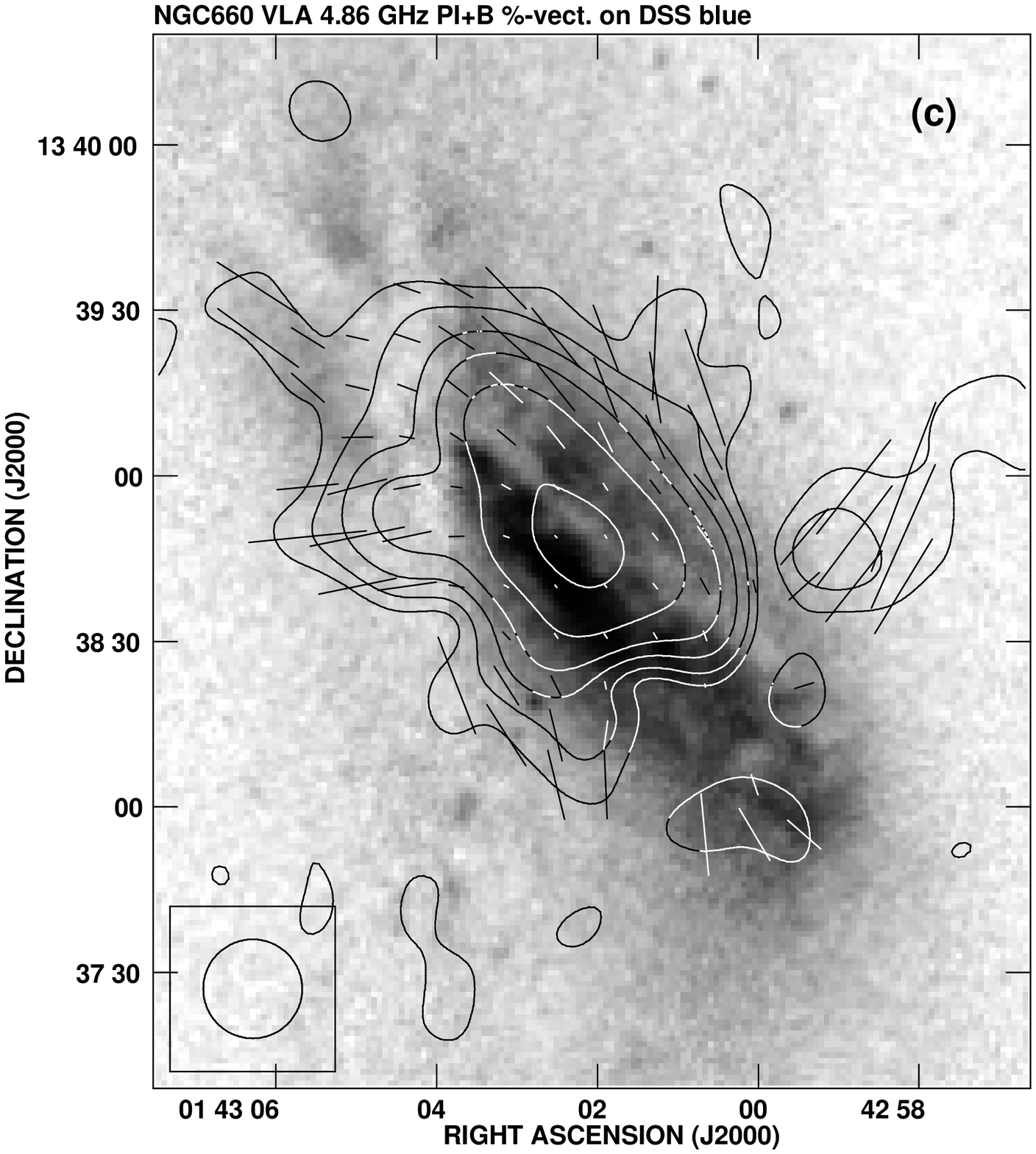}
\end{minipage}
\caption{a) The total power contours of NGC$\,$660 at 4.86$\,$GHz (natural weighting) superimposed on the DSS blue image. 
A large warped ring in the NS direction is visible. The contour levels are (-3, 3, 5, 8, 12, 20, 35, 80, 150, 250, 400, 700, 1500, 4000) $\times$
26.4 $\mu$Jy/beam. b) The total power contours and B-vectors of polarized intensity of the disk of NGC$\,$660 superimposed on DSS blue 
image. The contour levels are the same as for Fig. a). A vector of 10$"$ length corresponds to the polarized intensity of 100.0 $\mu$Jy/beam. 
c) The contours of polarized intensity and B-vectors of polarization degree of the disk of NGC$\,$660 at 4.86$\,$GHz (natural weighting) 
superimposed on the DSS blue image. The contour levels are (3, 5, 8, 12, 20, 35) $\times$ 13.4 $\mu$Jy/beam.
A vector of 10$"$ length corresponds to the polarization degree of 11\%. The maps resolution is 18$"$ $\times$ 18$"$ HPBW.
}
\label{f:n660tp}
%\end{center}
\end{figure*}

\subsection{The Taffy2}
\label{s:taffy2}

This pair of post-collision spiral galaxies (UGC\,813 and UGC\,816) is 
similar to the Taffy system and shows a radio bridge at 1.40\,GHz (Condon et 
al. \cite{condon02}). The suspected head-on collision was also a recent one that occurred 
about $5\times 10^7$\,yr ago. We call this system the Taffy2.

Our re-reduced data at 4.86\,GHz show that, besides the radio bridge, the regular field is 
discerned in the barred disk of UGC\,816 and resembles a typical spiral structure 
(Fig. \ref{f:taffy2tp}). The configuration of $\vec{B}$-vectors 
is not as clear in the second galaxy, which could be caused by its high inclination 
(72\degr). The orientation of magnetic field is almost perpendicular to the 
line joining the disks, which is somewhat puzzling, considering the opposite orientation of 
magnetic field in the Taffy system. 
The uncertainty of magnetic field orientations estimated from the signal-to-noise ratio 
of the polarized intensity is about $10\degr$. On the 
other hand, the foreground rotation measure of the Milky Way 
is about 10\,rad\,m$^{-2}$ for the Taffy coordinates and about $-50$\,rad\,m$^{-2}$ for 
the Taffy2 (Johnston-Hollitt et al. \cite{johnston04}). Thus, the uncertainties of the
magnetic field orientation due to the foreground RM are $2\degr$ and $11\degr$. 
Therefore, neither uncertainties in determination of orientations of B-vectors nor 
the foreground RM could cause such large differences in the observed 
patterns of magnetic fields.

Following from this, magnetic field orientations perpendicular to the interaction axis in the 
Taffy2 shows that this system is less taffy-like than the Taffy one. 
This may indicate less violent interaction in Taffy2, hence weaker distortions 
of magnetic fields out of its galactic disks. 
The rough estimation of kinetic energy involved in gravitational interactions are  
$4.9-14.0\times 10^{57}$\,erg and $3.2-4.6\times 10^{57}$\,erg 
for Taffy and Taffy2 galaxies, respectively (Lisenfeld \& Volk \cite{lisenfeld10}). 
The Taffy2 thus indeed gives a hint of weaker interaction.
Another possibility for different magnetic field orientations 
is a former collision in Taffy2 giving 
a longer time for galactic rotation to re-establish the global galactic magnetic field  
(especially in UGC\,816, cf. Fig. \ref{f:taffy2tp}). 
The differences in magnetic field configuration may also arise from projection effects, 
provided the magnetic field lines are not straight but are, e.g., helical in the bridges 
(M. Soida, private communication). Detailed MHD simulations are needed to 
evaluate these possibilities.

\subsection{NGC\,660}

NGC\,660 is usually recognized as a polar-ring galaxy with a large warped ring of 
gas and stars (Fig. \ref{f:n660tp}a). It  was suggested that its morphology 
results from a central collision and subsequent merging of two spiral galaxies of 
similar masses that happened several Gyrs ago (van Driel et al. \cite{driel95}). 
However, numerical simulations can indicate another scenario where the tidal interactions 
between two objects without a final merger caused accretion of  polar-ring material to NGC\,660 from 
the donor galaxy (Bournaud \& Combes \cite{bournaud03}).

Our map of the total radio intensity at 4.86\,GHz (Fig. \ref{f:n660tp}b) displays a smooth 
emission from a disk with a strong central component. The central source 
has a flux of $85^{+5}_{-8}$\,mJy, which agrees with its 
total flux at 1.41\,GHz and the spectral index $\alpha=0.6$ 
($S\propto \nu^{-\alpha}$) estimated by Condon et al. (\cite{condon82}). 
Surprisingly enough, the directions of $\vec{B}$-vectors 
form an undistorted `X-like' structure in the outer part of the disk 
and parallel to the disk in its inner part (Fig. \ref{f:n660tp}c). Such a configuration 
is typically observed in normal late-type spiral galaxies seen edge-on 
(e.g. Braun et al. \cite{braun10}). We suggest that in this case
the second evolutionary scenario with tidal accretion, rather than one involving 
a strong disk deformation 
(expected for the merger), is more likely for this system. The convolution of our maps
to the larger beam of $1\arcmin$ does not reveal any radio protrusion from the polar ring. This 
could be expected, because in either evolutionary scenario,  
the life-time of relativistic electrons at 4.86\,GHz (about $10^8$\,yr) is 
much shorter than the estimated age of the ring (about $10^9$\,yr; 
van Driel et al. \cite{driel95}).  

\subsection{Other interacting and merging systems}
\label{s:other}

{\em NGC\,4254} --- This is a weakly interacting spiral galaxy in the the Virgo cluster,
which has recently experienced a gravitational encounter to have perturbed its spiral arms 
and triggered a burst of star formation (Chy\.zy et al. \cite{chyzy07b}). 
The galaxy shows a polarized emission throughout the disk, particularly strong 
in its southern ridge, shifted downstream of a density wave with up to $13\,\mu$G as a regular 
magnetic field component and $20\,\mu$G in the total field (Chy\.zy \cite{chyzy08}). In 
the ridge, the dynamo-induced magnetic fields are modified by stretching and shearing forces, 
which are likely triggered off by tidal forces, which produced an anisotropic 
component of regular field and enhanced the polarized emission.  

{\em NGC\,4038/9} --- This well-known interacting pair of almost identical 
spiral galaxies (``the Antennae'') is probably just prior to the second encounter, which will 
lead to their merging in $2\times 10^8$\,yr time (Mihos et al. \cite{mihos93}; 
Kotarba et al. \cite{kotarba10}). The polarization properties of this system 
were studied in detail by Chy\.zy \& Beck (\cite{chyzy04}, see the Introduction). 
The strongest and almost random magnetic fields ($30\,\mu$G) are located in 
the overlap region of both the galactic disks. In the NE tidal tail, away 
from star-forming regions, magnetic field is highly coherent 
with a strong regular component of $10\,\mu$G, tracing gas shearing 
motions along the tail. The radio spectrum is much steeper in this region, 
indicating ageing of the CR electron population as they propagate from their 
sources in the star-forming regions. This system was chosen by Toomre (\cite{toomre77}) 
as the first step in his {\em merging sequence}. 

{\em NGC\,6621} --- This galaxy, together with NGC\,6622, is included in the Toomre 
sequence as an intermediate-stage merger. At 1.49\,GHz we see in our maps only 
the total power emission associated with the optically brightest part of the system. 

{\em NGC\,520} --- This system represents an intermediate stage of galaxy interactions 
according to Toomre (\cite{toomre77}). 
The radio emission at 1.4\,GHz is strong in the central part of the system and 
shows a protrusion to the NW along the optically bright ridge. 
According to Stanford \& Barcells (\cite{stanford91}), the system is at present 
$3\times 10^8$\,yr past the encounter and prior to the final coalescence.

{\em NGC\,3256} --- A kinematic study of this ULIRG system (English et al. \cite{english03}) 
indicates that it is currently experiencing a starburst just prior to the merging. The 
two tidal tails visible in optical images probably formed about $5\times 10^8$\,yr ago and 
the coalescence is estimated to take place in $\approx 2\times 10^8$\,yr. The 
radio data at 4.86\,GHz reveal only total emission peaked at the galactic centre,  
the site of intense infrared and X-ray emission (Brassington et al. \cite{brassington07}).

{\em Arp\,220} --- This is the closest ULIRG, the merger at the point of coalescence 
(Brassington et al. \cite{brassington07}). The radio emission at 4.86 and 1.43\.GHz is 
highly concentrated and coincident with the optical one. 

{\em NGC\,7252} --- This well-known merger, called ``The Atoms for Peace'', 
shows remarkable optical loops and tidal tails surrounding it 
(Dopita et al. \cite{dopita02}). This is the last system within the Toomre 
(\cite{toomre77}) sequence of interacting galaxies. The star formation is 
strongly concentrated towards the nucleus and has a counterpart in the radio 
emission at 4.86\,GHz. According to Mihos et al. (\cite{mihos93}), the SFR had 
dropped in this object by two thirds from the time of coalescence, 
estimated to occur $\approx 1$\,Gyr ago. 

{\em Arp\,222} --- A post-merger galaxy is at a slightly more
advanced stage of evolution than NGC\,7252 (Brassington et al. \cite{brassington07}).
The protrusions and plums observed in optical and X-ray emission indicate that the system has not 
relaxed to an elliptical galaxy yet. Our analysis of available VLA radio data of this system 
at 1.49\,GHz provides the total emission from the central part of the system.

{\em NGC\,1700} ---  A protoelliptical galaxy possesses a counter-rotating 
stellar core and other symptoms that it formed through merging of two 
spiral galaxies of similar mass (Brassington et al. \cite{brassington07}). The best 
age estimate for the coalescence event is about 3\,Gyr (Brown et al. 
\cite{brown00}). No radio emission of this galaxy was  detected in the NVSS survey, so 
only the upper limit of radio emission at 1.4\,GHz could be estimated 
(Table~\ref{t:notreduced}).

\subsection{Magnetic field strength and regularity}
\label{s:magnetic}

\begin{table}[t]
\caption{Magnetic field strength and regularity for interacting galaxies.}
\begin{center}
\begin{tabular}{lcccc}
\hline\hline
Name & B$_{tot}$ & B$_{reg}$ & Field \\
 & [$\mu$G] & [$\mu$G] & Regularity \\
\hline
NGC\,4254  & 15 $\pm$ 4 & 7 $\pm$ 2 & 0.48 $\pm$ 0.17 \\
NGC\,5426  & 11 $\pm$ 3 & 3 $\pm$ 1 & 0.32 $\pm$ 0.13 \\
NGC\,5427  & 13 $\pm$ 4 & 4 $\pm$ 2 & 0.33 $\pm$ 0.13 \\
NGC\,876   & 11 $\pm$ 3 & 2 $\pm$ 1 & 0.21 $\pm$ 0.09 \\
NGC\,877   & 15 $\pm$ 5 & 4 $\pm$ 2 & 0.24 $\pm$ 0.09 \\
NGC\,2207  & 16 $\pm$ 5 & 6 $\pm$ 2 & 0.39 $\pm$ 0.15 \\
IC\,2163   & 12 $\pm$ 4 & 4 $\pm$ 2 & 0.33 $\pm$ 0.15 \\
NGC\,6907  & 15 $\pm$ 4 & 3 $\pm$ 1 & 0.23 $\pm$ 0.10 \\
UGC\,12914 & 12 $\pm$ 4 & 3 $\pm$ 1 & 0.27 $\pm$ 0.11 \\
UGC\,12915 & 15 $\pm$ 4 & 2 $\pm$ 1 & 0.15 $\pm$ 0.07 \\
UGC\,813   & 13 $\pm$ 4 & 3 $\pm$ 1 & 0.21 $\pm$ 0.09 \\
UGC\,816   & 15 $\pm$ 5 & 3 $\pm$ 1 & 0.19 $\pm$ 0.09 \\
NGC\,660   & 16 $\pm$ 5 & 3 $\pm$ 1 & 0.18 $\pm$ 0.08 \\
NGC\,4038  & 18 $\pm$ 6 & 4 $\pm$ 2 & 0.24 $\pm$ 0.11 \\
NGC\,4039  & 12 $\pm$ 4 & 3 $\pm$ 2 & 0.23 $\pm$ 0.12 \\
NGC\,6621  & 13 $\pm$ 4 & N/A & N/A\\
NGC\,520   & 13 $\pm$ 5 & N/A & N/A\\
NGC\,3256  & 25 $\pm$ 8 & N/A & N/A\\
Arp\,220   & 27 $\pm$ 7 & N/A & N/A\\
NGC\,7252  & 12 $\pm$ 4 & N/A & N/A \\
Arp\,222   & 5  $\pm$ 2 & N/A & N/A\\
NGC\,1700  & $<$6 $\pm$ 2 & N/A & N/A\\
\hline
\end{tabular}
\end{center}
\label{t:b}
\end{table}

The radio emission detected by us for interacting galaxies is composed 
of nonthermal (synchrotron) and thermal (free-free) radiation. The 
nonthermal emission and its polarized component can be used to obtain 
total and regular (ordered) magnetic field strengths 
(Beck \& Krause \cite{beck05}, see Appendix \ref{s:formula} for details). 
A typical equipartition energy condition between cosmic 
rays and magnetic fields can be applied, especially as spatial resolution
of the the analysed radio observations are as large as hundreds of parsecs. 
For all galaxies the total and polarized radio intensities were integrated 
over the galactic disks as outlined by the optical emission visible in optical B-images. 
We assumed a typical unprojected pathlength through the synchrotron 
emitting region of 1\,kpc. However, for NGC\,1700, which is perfectly an 
edge-on object (Table \ref{t:sample}) without detectable radio emission, 
we adopted the synchrotron pathlength $L$ of 2\,kpc (see Appendix \ref{s:formula}).

The nonthermal spectral indexes and thermal fractions for NGC\,2207/IC\,2163, 
NGC\,660, NGC\,4038/9, and NGC\,4254 were taken from the separation 
of radio emission components by Niklas et al. (\cite{niklas97}), 
Chy\.zy \& Beck (\cite{chyzy04}), and Chy\.zy et al. (\cite{chyzy07b}).
For other objects, a typical nonthermal spectral index of 0.8 was used 
(Niklas et al. \cite{niklas97}), and the radio thermal emission was estimated 
from the FIR  data (Surace et al. \cite{surace04}; Sanders et al. 
\cite{sanders03}; Brassington et al. \cite{brassington07}; 
Bushouse et al. \cite{bushouse88}; Moshir et al. \cite{moshir90}) in the following way. 
First, we derived global SFRs from 
galactic infrared luminosities at 60 and $100\,\mu$m, according to the 
empirical SFR indicator derived by Kennicutt (\cite{kennicutt98}). 
They range from less than $1$\,M$_{\sun}$\,yr$^{-1}$ 
(NGC\,5426, ARP\,222, NGC\,1700) up to $150$\,M$_{\sun}$\,yr$^{-1}$ (ARP\,220), 
with uncertainties less than 25\%. Next, we converted the obtained SFRs 
to the predicted radio thermal emission, according to the modelling of free-free 
radiation by Caplan \& Deharveng (\cite{caplan86}), see also Chy\.zy et al. (\cite{chyzy07b}). 
The estimated thermal fractions from this method are about 0.10 at 
4.86\,GHz with uncertainties of about $0.02$. Synchrotron emission of 
galaxies was obtained by subtraction of their radio thermal components from the 
total radio emission. 

As a consistency test of the applied approaches to calculating galactic 
magnetic field strengths, we compared the two methods of estimating of thermal 
emission: from the FIR SFR indicator and from the separation
of thermal and non-thermal radio emission. 
For example for a weakly interacting galaxy NGC\,4254 for which magnetic properties
are well known (Chy\.zy \cite{chyzy08}), the first method gives 
$B_{tot}=14\pm 4\,\mu$G, $B_{reg}=6\pm 2\,\mu$G. From the second method we have 
$B_{tot}=15\pm 4\,\mu$G, $B_{reg}=7\pm 2\,\mu$G.
Therefore, both methods give consistent results that are well within 
estimated uncertainties. We also note that the 
magnetic field strength only weakly depends on the thermal fraction (Eq. 
\ref{e:B} in Appendix \ref{s:formula}).

The derived strengths of the total and regular magnetic fields for all our 
interacting galaxies are given in Table ~\ref{t:b}. 
The presented uncertainties include uncertainties of estimated 
non-thermal radio brightness, a factor of 2 uncertainties in adopted proton-to-electron 
density ratio and pathlength $L$, and the uncertainty of 0.1 of applied nonthermal spectral 
indexes. The mean values of total magnetic field strength $B_{tot}$ obtained for 
the interacting objects range from $5-6\,\mu$G for post-mergers Arp\,222 
and NGC\,1700 to $25-27\,\mu$G for ongoing mergers NGC\,3256 and Arp\,220 
(Table~\ref{t:b}). This spread may result from the diversity in properties 
of individual galaxies, as well as from the particular 
stages of their gravitational interactions (see the next Section). 
The averaged total magnetic field strength for the whole sample 
is $14\pm5\,\mu$G. Excluding the two lowest values 
for the post-mergers (Arp\,222 and NGC\,1700), which much differ from the rest of the sample, 
the  mean strength rises to $15\pm4\,\mu$G. Both the means are more than the typical average total field 
strength of $9\,\mu$G estimated for 74 bright spiral galaxies (Niklas 
\cite{niklas95}, Beck \& Krause \cite{beck05}). As the previous studies of magnetic 
fields have involved all morphological types of galaxies, the cited mean is likely 
increased by including of interacting objects. For instance, the average 
strength of total magnetic field for typical large spirals in our 
neighbourhood (the Galaxy and M\,31)  
is $5-7\,\mu$G, which is significantly lower than for the interacting objects in our 
sample. 

We also calculated the strength of regular magnetic field 
$B_{reg}$ and the field regularity $B_{reg}/B_{ran}$, which 
is a useful measure of relative production of regular and random magnetic 
fields (Table~\ref{t:b}). In contrast to the total magnetic field, the strength of 
regular component is similar to those in the non-interacting spirals, which leads to a mean of 
$2.5-3.5\,\mu$G. The enhanced total field with relatively normal 
regular component results in a small field regularity ($0.27\pm0.09$) in 
the interacting objects.

The local magnetic fields are strongest in the centres of galaxies and reach 
$48\,\mu$G in Arp\,220 and $53\,\mu$G in NGC\,3256. Similar values were 
also observed in the centres of most starbursting galaxies, as in M\,82 (Klein et al. 
\cite{klein88}). The regular field peaks in the regions of weak total 
radio intensities, far off the regions of intense star formation. 
For example, in the bridges of Taffy and Taffy2 galaxies, $B_{reg}$ reaches 
$11\pm 3\,\mu$G. This value was obtained for the nonthermal
spectral index of $1.3\pm0.1$ approximated from the spectral 
index maps of Condon et al. (\cite{condon93}) and Condon et al. (\cite{condon02}). 
Furthermore, we adopted a synchrotron pathlength of about 15\,kpc corresponding 
to the width of the observed radio bridges (cf. Figs. \ref{f:taffytp} and 
\ref{f:taffy2tp}). A similar strength of regular field was observed at the base of 
the eastern tidal tail in the Antennae system (Chy\.zy \& Beck \cite{chyzy04}). 
Because in all such regions the total 
field is also strong, i.e. $16\pm4\,\mu$G in the bridges 
of the Taffy and Taffy2 systems, the magnetic field can affect the local gas dynamics. Actually, in the bridge of 
the Taffy system, the magnetic energy of $10.2 \times10^{-12}$\,erg\,cm$^{-3}$ 
is much larger than the energy of thermal gas, which is 
$1.4 \times10^{-12}$\,erg\,cm$^{-3}$ (as estimated from the radio thermal 
emission at 4.86\,GHz, assuming $10^4$\,K gas temperature, and the volume filling 
factor of 0.5). The  turbulent velocity of \ion{H}{i} gas of about 30\,km\,s$^{-1}$ suggested 
by Condon et al. (\cite{condon93}) results in the turbulent energy of ISM of 
$6.6\times10^{-12}$\,erg\,cm$^{-3}$, again lower than the magnetic energy. 
Although the kinetic energy of \ion{H}{i} blobs orbiting 
the system is about two orders of magnitude higher, locally magnetic fields can still affect 
the gas phase, directing the gas motion along the magnetic field lines. 
Similar conclusions are valid for the bridge between the Taffy2 galaxies.

\section{Statistical analysis}
\label{s:discussion}

It is believed that gravitational interactions between galaxies can induce star formation 
(see Struck \cite{struck06} for a review).
On the other hand, star-forming activity is one of the main factors in determining dynamo 
processes and regulating random and regular magnetic fields in galaxies (e.g. Chy\.zy \cite{chyzy08}).
We have shown that the interacting galaxies reveal various radio structures and
show a range of magnetic field strength (Table \ref{t:b}).
Therefore, we investigate how, statistically, magnetic fields and radio morphologies 
are connected to the SFR in the interacting objects. Having optical sizes of galaxies (corrected for 
galactic extinction and inclination, from HyperLeda), we are able
to build similar relations for the surface densities of the SFRs ($\Sigma$SFR).

In order to relate the properties of radio emission and magnetic fields to
different stages of gravitational interactions we introduce ``the interaction stage
parameter", $IS$, which is simply a number indicating the particular stage 
of interaction and galaxy morphological distortion.
Toomre (\cite{toomre77}) arranged 11 ongoing mergers according to the 
rising strength of observed and modelled tidal interactions 
(see Introduction). We assigned them consecutive 
$IS$ values from 1 to 11 (Table \ref{t:sample}). The values 1-9 correspond 
to pre-merger stages. $IS=1$ describes systems after the first galaxy encounter 
and with significant disk distortions (e.g the Antennae and Taffy systems). 
Galaxies with stronger interactions, evolving gradually to coalescence, 
are NGC\,6621/2 ($IS=5$), NGC\,520 ($IS=7$), and NGC\,3256 ($IS=9$), all displaying 
strong and concentrated radio emission associated with the optically brightest
region in the system. The nuclear coalescence corresponds to $IS=10$ (Arp\,220), 
and the young merger remnant to $IS=11$ (e.g. NGC\,7252).
We supplemented the Toomre sequence with weakly interacting galaxies 
from our sample: we assign $IS=-1$ to objects before the first encounter 
(e.g. NGC\,876/7) and $IS=0$ to galaxies close to the (first) encounter 
(e.g. NGC\,6907/8) when the tidal tails are launched 
(Sect. \ref{s:results}; Table \ref{t:sample}).  
The other side of the sequence we expand with age-advanced mergers 
but still not relaxed as for Arp\,222 ($IS=12$, cf. Brassington et al. 
\cite{brassington07}). Our last object NGC\,1700 is a pre-elliptical system 
($IS=13$) that nevertheless shows evidence that it was formed from spiral 
galaxies (Brassington et al. (\cite{brassington07}).
It is clear that the attributed values of $IS$ parameter may be affected by 
different interaction geometries or properties of individual galaxies. 
However, to the first approximation, they provide a rough indicator of the evolutionary stage and ``strength'' of galaxy interaction.

\subsection{Evolution of magnetic fields}
\label{s:evolution}

\begin{figure}[t]
\centering
\includegraphics[width=9.0cm]{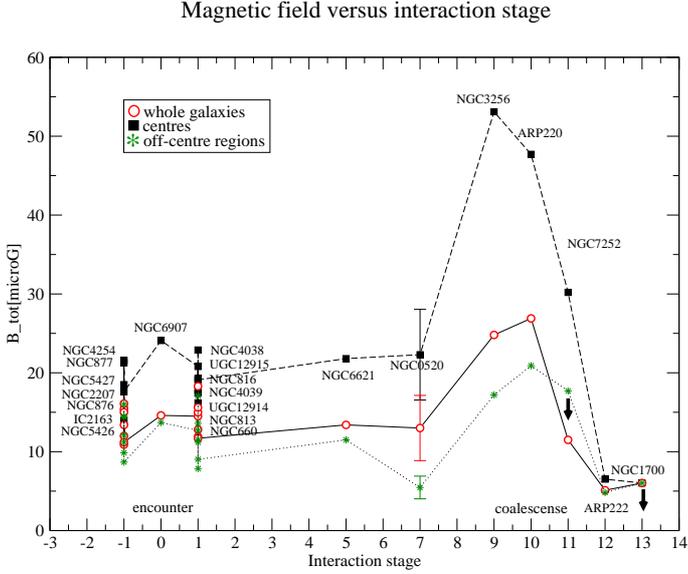}
\caption{The evolution of magnetic field strength in interacting galaxies. 
Mean magnetic field strengths are given for the whole galaxies (circles), 
centres (squares), and off-centre regions (asterisks). Arrows indicate upper limits of 
the field strength. Two special stages of interaction are denoted: the first 
galaxy encounter and the coalescence of 
the merger nuclei. The error bar for NGC\,520 is shown.}
\label{f:evolution}
\end{figure}

We explore the evolution of magnetic fields through the interaction stages of galaxies 
using the introduced $IS$ parameter (Fig. \ref{f:evolution}). For the weakly interacting 
galaxies, the strength of magnetic field is almost constant ($10-15\,\mu$G). 
However, we show for the first time that there is a distinct trend toward the increasing 
strength of magnetic field as the interaction advances towards the nuclear coalescence 
(up to $25\mu$G for Arp\,220; $IS=10$). A completely opposite trend is found at 
the later stages, i. e. for post-mergers, 
for which the strength of magnetic field decreases in a systematic manner. 
Neither radio emission nor magnetic fields are detected for the last galaxy in 
our sequence, which is a proto-elliptical object (NGC\,1700).
Similar trends for the whole galaxies are observed when magnetic field 
strength is calculated only for the central regions of objects (of about 
6.5\,kpc size) or outside of them in the galactic disks (Fig. \ref{f:evolution}).

The diverse trend found in magnetic field evolution seems to be 
connected with the star-forming activity in the interacting systems.
This view is supported by the results of Georgakakis et al. (\cite{georgakakis00}), 
who show for a similar merger sequence that the FIR luminosity to molecular hydrogen 
mass ratio (the star formation efficiency) clearly increases close to the nuclear 
coalescence and 
declines again after merging. The strongly interacting and merging galaxies 
typically have star formation efficiency about an order of
magnitude higher than that of the isolated and weakly interacting galaxies. 
Georgakakis et al. also show that, during the merging, the enhancement of star formation is moderate
within the disk, while strong in the nucleus. In accordance with that, we observe  
that the major enhancement of magnetic energy in galaxies in our sample also occurs 
in galactic centres. The field strength increases here by more than 170\%, whereas within the 
disk the rise is less than 100\% (Fig. \ref{f:evolution}).
A rise in nuclear luminosities has also been found in optical (Laine et al. \cite{laine03})
and NIR observations with the HST (Rossa et al. \cite{rossa07}).

The \ion{H}{i} study of the Toomre merging sequence by Hibbard \& van Gorkom 
(\cite{hibbard96}) indicated that the hydrogen gas is pushed out from galactic 
disks into tidal features as the merging process advances. Such features are 
not traced by synchrotron emission on our radio continuum maps. 
Moreover, the large velocity dispersion estimated from \ion{H}{i} measurements
in weakly interacting galaxies (e.g. NGC\,2207, Sect. \ref{s:2207}), likely caused by 
tidally induced turbulence or large-scale gas motions, are not associated with 
the observed radio features.
Therefore, the star-forming activity must 
be the main energy source for the radio (synchrotron) emission and magnetic fields 
in merging galaxies.
The revealed evolution of magnetic fields agrees with the scenario of 
formation of ellipticals from mergers (Toomre \& Toomre \cite{toomre72}), 
because there is no detection of synchrotron emission from the 
elliptical galaxies to date, apart from the AGN phenomena (Sect. \ref{s:intro}). 

The evolution of magnetic fields through the interaction stages differs significantly from what is observed in the X-rays. We do not find any enhancement of magnetic field 
long after the nuclear coalescence, which is observed in the X-ray emission in 
the merging sequence explored by Brassington et al. (\cite{brassington07}). 
Therefore, it seems that unlike with X-rays, there is no process connected to 
Type I supernova that could re-create the magnetic field in the merger remnants, as was proposed to account for the X-ray observations (Sect. \ref{s:intro}). 
Magnetic fields and cosmic rays require powerful Type II supernova explosions 
for their generation (Beck et al. \cite{beck96}). 
Therefore, our findings suggest that the main production of magnetic fields in colliding 
galaxies must terminate close to the coalescence, after which 
magnetic field diffuses or is kept at the level of the turbulent energy of ISM. 
In either case, for the radio frequencies considered in this paper, the synchrotron 
emission should fade in proto-elliptical (merged) galaxies after about $10^8$\,yr if there are no 
sources of CR energization. The fading synchrotron emission in post-merger systems
is too weak to be detected using the currently available radiointerferometric data. 
New radio instrumentation operating at extremely low frequencies such as LOFAR 
could possibly detect the faint synchrotron emission from the old population of CR 
electrons, revealing such ``relic'' magnetic fields.

\subsection{Magnetic fields versus star formation rate}
\label{s:sfr}

\begin{figure}[t]
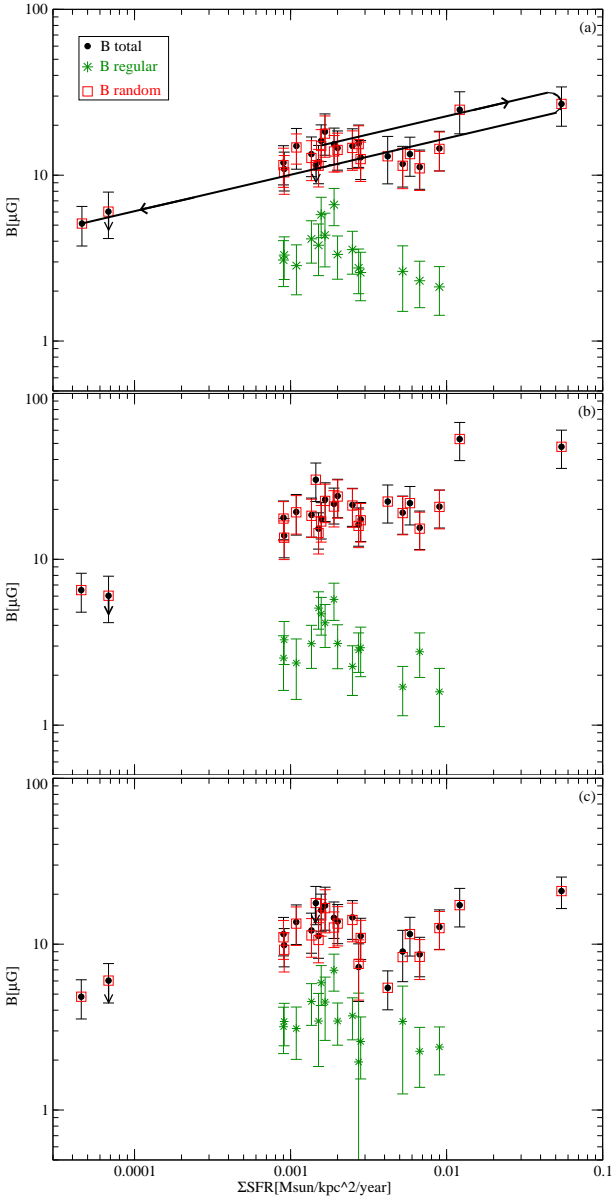

\centering
\includegraphics[width=8.0cm, trim=0 42 0 0]{16092fg13a.eps}
\includegraphics[width=8.0cm, clip=true, trim=0 41 0 7]{16092fg13b.eps}
\includegraphics[width=8.0cm, clip=true, trim=0 0 0 7]{16092fg13c.eps}
\caption{Total magnetic field strengths (circles) versus local SFRs for the whole interacting 
galaxies a), their centres b), and off-centre regions c). The random and regular 
components of magnetic fields are also provided (squares and asterisks, respectively). An
expected evolutionary track of interacting systems is also indicated.
}
\label{f:fields}
\end{figure}
In Fig. \ref{f:fields}a we show total magnetic field strength related to 
the $\Sigma$SFR for 22 galaxies for which radio emission could be determined (except for 
NGC\,6908 and NGC\,6622, see Sect. \ref{s:results}). We also plotted mean magnetic field strengths in the 
galactic centres (Fig. \ref{f:fields}b) and off-centre regions (Fig. \ref{f:fields}c).
We found that the statistical relation for the entire galaxies is weak (the Spearman correlation 
coefficient $\rho=0.49$) with an index of fitted power law  $\alpha=0.20\pm 0.02$.
The dependence is stronger in galactic centres ($0.27\pm 0.03$) and weaker outside them 
($0.14\pm 0.03$). We notice that, within a wide range of the  $\Sigma$SFR 
of $10^{-3}-10^{-2}\,\rm{M}_{\sun}\,\rm{yr}^{-1}\,\rm{kpc}^{-2}$, the magnetic field strength is actually spread over $10-20\,\mu$G without 
any discernible tendency. In fact, the determined slopes  
are established mainly thanks to the galaxies of weakest and of strongest 
$\Sigma$SFR, which are all members of particularly strongly interacting systems ($IS\ge9$).

We also constructed diagrams similar to Fig. \ref{f:fields} but with the global value of SFR. 
The slopes and correlations are statistically similar (e.g. $\alpha=0.21\pm 0.02$, $\rho=0.63$
for the entire galaxies). This therefore seems to be a general
phenomenon that magnetic fields are only weakly dependent 
on galactic star-forming activity, showing a wide spread of estimated 
strengths. The wider the range of considered SFRs,
the better defined and stronger the influence on magnetic fields.
This result can explain the former findings of Hummel (\cite{hummel81}) 
that the radio continuum properties of interacting galaxies differ from those for the 
isolated objects only in the central parts of galaxies, and not in their disks. 
His analysis ignores, however, all the advanced mergers that we included in our sample (NGC\,3256, 
Arp\,220, Arp\,222, NGC\,1700). Similarly, as shown by Kennicutt et al. (\cite{kennicutt87}), only particularly strongly interacting galaxies show a distinct enhancement 
in star formation, as appraised from the H$\alpha$ and FIR emission.

No mutual relation is found between $\vec{B}$ and SFR (Fig. \ref{f:fields}), 
as observed for $\vec{B}$ and $IS$ (Fig. \ref{f:evolution}). It is because 
both the advanced mergers and the weakly interacting galaxies 
have low SFRs, and they are undistinguishable in Fig. \ref{f:fields}. 
Only when the magnetic field is set against the stage of interaction (Fig. \ref{f:evolution}), is the full evolution of $\vec{B}$ revealed. We plot in Fig. \ref{f:fields} an
evolutionary track of exemplary interacting system as based on our results. 

There is a hint of negative correlation between the regular component of 
magnetic field and the $\Sigma$SFR for entire galaxies, $\rho=-0.55$, and the 
power-law fit is $-0.25\pm0.07$ (Fig. \ref{f:fields}a). The same trend was more 
distinct for the magnetic field regularity of individual regions of the weakly interacting 
galaxy NGC\,4254 (Chy\.zy et al. \cite{chyzy07b}). Magnetic field tangling in active star-forming 
regions and production of random magnetic fields by the local-scale dynamo process 
(Brandenburg \& Subramanian \cite{brandenburg05}) result in a lower field regularity. This may also be 
the case over the galactic scales in the analysed interacting objects.

\subsection{Asymmetries}
\label{s:asymmetries}

\begin{figure}[t]
\centering
\includegraphics[width=8.0cm]{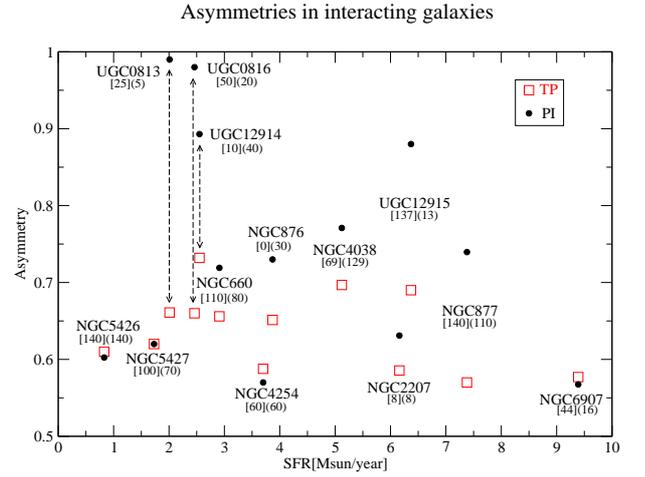}
\caption{Asymmetry parameters for total ($A_T$; squares) and polarized radio emission 
(circles; $A_P$) versus global SFR for interacting galaxies. The values of 
offset angles [$\phi_T$]( $\phi_P$) for asymmetry axes in total and 
polarized emission are given in parentheses. Typical uncertainties of
asymmetry parameters are less than $5\%$.
}
\label{f:asym}
\end{figure}

Gravitational interactions are able to severely distort galactic disks (Sect. 
\ref{s:intro}). Here, we investigate how they modify distributions of radio 
total and polarized emission at 4.86\,GHz. First, we divided each galaxy into 
two parts by defining an axis in the N-S direction that crosses the galactic 
centre. We integrated the total radio intensities over the obtained galactic 
halves ($T_1$, $T_2$) and determined the ratio $a_T=T_1/(T_1+T_2)$, with a convention of $a_T\ge0.5$.
Next we rotated the radio map by $30\degr$ around the galactic optical centre 
and  determined $a_T$ once again. The whole procedure was repeated with the step in the rotation of 
$30\degr$, until the map involved all orientations (six positions). Finally, we 
determined the galaxy radio asymmetry $A_T$ as the highest of all $a_T$ values. 
We also calculated the difference $\phi_T$ between the original position angle 
of the major axis of the galaxy (Table \ref{t:sample}) and the position angle of 
the major axis after rotation, which gave the asymmetry $A_T$.
A similar procedure was applied to the radio polarization map, 
which likewise yielded the polarization asymmetry $A_P$ and the offset angle of 
the polarization asymmetry axis $\phi_P$. To estimate the accuracy of 
obtained values of asymmetry parameters, we rotated the axes which gave asymmetries 
$A_T$ and $A_P$ by $\pm 15\degr$. The uncertainties approximated as deviations from 
the original asymmetry parameters are small, typically less than $5\%$. 
All the above calculations were 
performed for those interacting galaxies for which total and polarized radio 
emissions were resolved and not too much affected by overlapping with 
a companion galaxy (Fig. \ref{f:asym}). 

\begin{figure}[t]
\centering
\includegraphics[width=8.0cm]{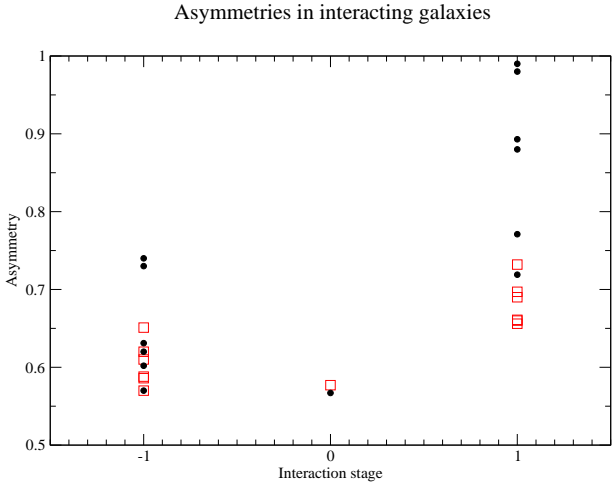}
\caption{Asymmetry parameters for total ($A_T$; squares) and polarized radio emission 
(circles; $A_P$) versus the interaction stage $IS$. Typical uncertainties of
asymmetry parameters are less than $5\%$.
}
\label{f:asym_merger}
\end{figure}

The obtained wide range of values of offset angles ($\phi_T$, $\phi_P$, 
presented in Fig. \ref{f:asym}) indicates 
that the asymmetries in total and polarized emission are not correlated with the 
apparent orientations of galactic disks. However, offset angles are within $60\degr$ 
from each other. The observed asymmetries are therefore not 
caused by projection effects and must be physically connected with galaxy 
morphologies as modified by the interaction process.
 
Among the analysed galaxies there is a trend that the asymmetry in distribution 
of polarized emission (mean of 0.75) is stronger than in the total radio intensity 
(mean of 0.64; Fig. \ref{f:asym}). The ordered magnetic 
field is therefore much more sensitive to morphological distortions than the total 
field. Even in the weakly interacting galaxies (e.g. NGC\,876/7), the polarized 
emission, as well as the orientation of magnetic field $\vec{B}$-vectors (cf. Sect. 
\ref{s:results}), are sensitive to global galactic distortions and anomalous 
gas flows. We also notice that the higher the distortion in the total emission 
($A_T$), the greater is the difference between the asymmetry in total and polarized 
emission ($A_P$). 

There is no obvious dependence of the strength of asymmetries on global SFRs for galaxies ($\rho=-0.35$; $-0.15$ 
for total and polarized radio emission, see also Fig. \ref{f:asym}). 
We also constructed a diagram similar to Fig. \ref{f:asym} but with 
the $\Sigma$SFR instead of global SFR. Also in this case the results
are similar ($\rho=-0.03$, $+0.12$). It seems that differences in the gas
content and mass of individual galaxies may blur 
the effect of possible enhancement of the SFR or $\Sigma$SFR by tidal interactions.

However, asymmetries are different before and after the first galactic 
encounter (Fig. \ref{f:asym_merger}). 
For all the weakly interacting objects ($IS=-1$; Table \ref{t:sample}), 
the asymmetries are small with the mean values $A_T=0.60\pm0.03$ and $A_P=0.65\pm0.07$. 
For all the objects after the 
first encounter ($IS=+1$), the asymmetries are systematically stronger with their means 
$A_T=0.68\pm0.03$ and $A_P=0.87\pm0.11$. In polarization they attain the highest measured 
values ($A_P>0.85$ for Taffy, Taffy2, and NGC\,4038). 
The exception from this behaviour is the galaxy NGC\,6907 ($IS=0$; 
Fig. \ref{f:asym_merger}), which is close 
to the moment of collision with NGC\,6908 (Sect. \ref{s:results}). However, 
otherwise than in the other galaxy pairs, this system is likely a minor merger and apparently 
the low-mass intruder (NGC\,6908) is not able to distort a much more massive galaxy. 
We thus conclude that the asymmetries visible in the radio emission in general only weakly depend 
on the global or local SFRs, while they increase (especially in the polarization) with 
advancing of interaction. 
As a result, the asymmetry in the radio polarized emission could be yet another indicator 
of an ongoing merging process.

\subsection{Radio-infrared relation}
\label{s:radiofir}

\begin{figure}[t]
\centering
\includegraphics[width=8.6cm]{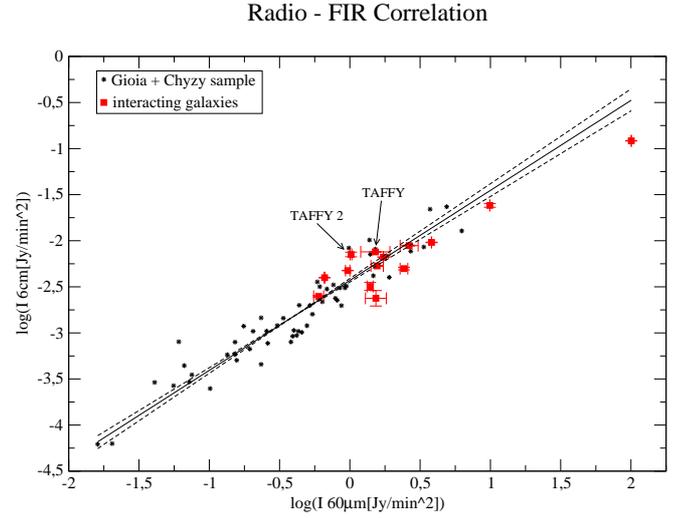}
\caption{The radio-FIR correlation diagram for interacting (squares)
and normal galaxies of different Hubble types (asterisks).
The surface brightness at 4.86\,GHz and at 60$\mu$m
(Jy/$\sq\arcmin$) are used (see text for details). 
The solid line is an orthogonal fit 
to normal galaxies, and dashed lines represent simple X vs. Y and Y vs. X 
regressions. Error bars for interacting galaxies are marked.
}
\label{f:radiofir}
\end{figure}

The process of tidal interaction and induced star formation may cause 
an enhancement in the infrared luminosity. This may lead to formation of 
strong infrared emitters as LIRG or ULIRG objects that are believed to be 
associated with colliding galaxies (Sect. \ref{s:intro}). In order to 
investigate whether radio emission also is correspondingly enhanced in such cases, 
we constructed a radio-FIR relation for our interacting systems and 
compared it to the one determined for a sample of normal galaxies of different 
Hubble types (Fig. \ref{f:radiofir}). We used the radio surface brightness of 
interacting objects estimated from our maps at 4.86 GHz and the infrared 
surface brightness at $60\,\mu$m derived from Surace et al. (\cite{surace04}), Sanders et al. (\cite{sanders03}), 
Bushouse et al. (\cite{bushouse88}), and Moshir et al. (\cite{moshir90}). For galaxies with only 1.4\,GHz flux available we scaled it to 4.86\,GHz assuming a typical spectral index of 0.8. 
We did not include two of the weakest radio objects Arp\,222 and NGC\,1700 for which $60\,\mu$m 
infrared data are unavailable. Typical uncertainties of obtained values 
are indicated in Fig. \ref{f:radiofir} and were derived from uncertainties 
of radio fluxes (Tables \ref{t:reduced} and \ref{t:notreduced}) 
and $60\,\mu$m fluxes (given in the mentioned FIR catalogues). 
For the sample of normal galaxies radio data at 4.85\,GHz were taken from 
Gioia et al. (\cite{gioia82}) and Chy\.zy et al. (\cite{chyzy07a}), whereas 
$60\,\mu$m data came from Helou \& Walker (\cite{helou95}) and 
Moshir et al. (\cite{moshir90}).

Our orthogonal fit to the radio-FIR relation for the sample of normal galaxies 
gives a power-law index of $0.98\pm 0.04$. The interacting galaxies follow 
this relation (Fig. \ref{f:radiofir}) and are located in the group of 
the strongest radio and FIR emitters. We note, however, that the brightest 
galaxy in our sample (Arp\,220) is shifted slightly down from the main trend
indicating possible flattening of the relation for relatively young mergers. A more 
numerous sample is needed to confirm this hint.
Recently we have shown that low-mass dwarf galaxies, which are by orders 
of magnitude weaker optical and radio emitters than the interacting objects, 
also follow a similar power-law relation, with a slope 
of $0.91\pm 0.08$ obtained for 2.64\,GHz radio data  
(Chy\.zy et al. \cite{chyzy11}).

In fact, the tidally induced intense star formation and morphological distortion 
in the studied interacting galaxies do not seem to change 
processes responsible for radio-FIR correlation.
This is quite interesting, as some of those systems are in the process of evolving 
and transformation towards S0 or elliptical merger remnants. 
The large-scale shocks, galactic large-scale, or small-scale dynamos, 
and processes mutually transforming magnetic field components 
(e.g. by compression and shearing, Sect.\,\ref{s:sfr}) therefore result in the production of 
magnetic energy and cosmic rays that are still well-balanced 
by production of thermal energy and dust radiation. 

Interestingly enough, according to our study, the deviations of the Taffy and Taffy2 galaxies 
from the main trend of the radio-FIR relation at 4.86\,GHz are small, 0.3 dex, and 
comparable to other objects from our sample (Fig. \ref{f:radiofir}). However, 
according to Condon et al. (\cite{condon93}; \cite{condon02}), these galaxies are twice 
as radio-loud as they ought to be if following the radio-FIR relation constructed 
at 1.4\,GHz. The difference between 4.86 and 1.4\,GHz is probably 
due to a more extended emission in the bridge at 1.4\,GHz and the steep 
radio spectral index (up to 1.3-1.4). The reason for such 
strong radio emission at 1.4\,GHz is not clear. According to Condon, 
the observed radio excess may come from the CRs electrons leaked from the 
galactic disks and trapped in enhanced (stretched) magnetic fields in the bridge 
between galaxies. 

However, Lisenfeld \& Voelk (\cite{lisenfeld10}) 
argue that in the Taffy systems, the kinetic energy of colliding gas probably brought about 
large-scale shock waves that accelerated CRs, as well as enhanced radio emission in 
the bridges. As this is not related to star formation, a deviation from the 
radio-FIR relation results. They assume in their modelling that the shocks in 
the Taffy disks correspond to the one in the middle-age supernova remnants and 
therefore could be described by the Sedov phase. However, according to our 
analysis of polarization data in the Taffy system (Sect. \ref{s:intro}), the magnetic field in the bridge is well-ordered and aligned with the direction of gas 
flows between disks. Thus the field configuration clearly resembles a very young 
supernova with magnetic field vectors aligned with gaseous outflows, well 
before the Sedov phase. As a result, the configuration 
of ordered magnetic fields indicates that the physics involved in 
explaining the Taffy bridge might not be applicable here. Alternatively,  
we deal here with some other processes of field ordering and compression, which 
work differently than in the supernova remnant. 
Sensitive observations of gamma rays could independently assess the CR population, 
indicate their origin, and account for the strong radio emission in the bridge.
In the Taffy2 system situation is different because magnetic field is not aligned 
with the bridge (Fig. \ref{f:taffy2tp}). This may indicate a better correspondence with the 
Lisenfeld \& Voelk model.

\subsection{The impact of merger's magnetic fields}
\label{s:impact}

Gravitational forces could play an important role in 
generation of magnetic field in the early-universe. 
If seed magnetic fields were produced by, e.g., Biermann battery 
or Weibel instability then small-scale dynamo during gravitational 
collapse of galaxies and accretion driven turbulence could strongly 
enhance the seed field (Schleicher et al. \cite{schleicher10}). Further 
evolution of galaxies could build up stronger, galactic-scale magnetic fields 
by small and large scale dynamos as discussed by Arshakian et al. (\cite{arshakian09}).
Gravitational forces during galaxy interactions and galaxy mergers could also play an important 
role in the magnetization of IGM.
Although the strongly interacting objects represent just a small fraction of galaxies in the local Universe, 
$<300$ interacting objects within 50\,Mpc distance according to the Vorontsov-Velyaminov et al.
(\cite{vorontsov01}) catalogue, they may have dominated the epoch of massive galaxy formation in the early Universe 
(Conselice \cite{conselice06}). Such galaxies are able to redistribute magnetic fields in IGM 
along tidal tails or bridges, as we in fact observe, e.g., in the Antennae 
and Taffy galaxies (Sect. \ref{s:results}). 
The lower limit on IGM magnetic fields of $5\times10^{-15}\,\mu$G was constrained 
by Tavecchio et al. (\cite{tavecchio10}) based on analysis of TeV and GeV emission from blazars.
Recent analysis of FERMI-LAT observations 
of blazars and the absence of their TEV envelopes can suggest that extragalactic magnetic 
fields must fill IGM with a large volume factor of $\approx60\%$ (Dolag et al. 
\cite{dolag10}). If such magnetic fields are not primordial, then they must have been 
redistributed from local sources, like colliding galaxies. 

To address the idea of magnetization of IGM by interacting 
objects, we approximate the zone around these galaxies that might be filled with 
magnetic fields simply by the largest extent of their observed \ion{H}{i} envelopes. The neutral 
hydrogen of colliding galaxies often reveals large protrusions that are not optically visible (Hibbard \& van Gorkom \cite{hibbard96}). If such gas is also accompanied by its partly 
ionized component, then the frozen magnetic field can be advected 
with gas flows and brought out from galactic disks. According to the collected literature data 
(Table \ref{t:sample}), we find that such outflows are up to about 200\,kpc in size for the 
most disrupted objects in our sample (e.g. Arp\,220). This value is compatible with the results 
of numerical simulations of high-redshift mergers (Bournaud et al. \cite{bournaud10}). The tidal interactions (estimated to operate within about 200\,kpc) are therefore not sufficient to seed 
vast volumes of IGM with magnetic fields, as required by the Dolag et al. conclusion. Either unknown 
transport processes redistribute magnetic fields 
of distant mergers out of expected \ion{H}{i} envelopes, or the proposed magnetization 
of space has a different origin. Even so, interactions can efficiently magnetize the merger surroundings. 
This could explain relatively strong magnetic fields ($\approx 10\,\mu$G) found 
around host galaxies of quasars with Mg II absorption systems up to redshift $z=1.3$ (Bernet et al. 
\cite{bernet08}).
Chy\.zy et al. (\cite{chyzy11}) suggest that massive galaxies in the redshift $z$=3-5 
could have magnetized IGM by supernova explosions and galactic winds to the extent 
of $\approx200-500$\,kpc. The process of strong gravitational interactions and associated 
magnetized outflows thus exerts a similar impact on IGM. 
New observational methods, such as a rotation measure synthesis, 
applied to background sources shining through merger remnants, could 
reveal such magnetized cocoons around mergers. New instrumentation such as the EVLA, 
LOFAR, and SKA could be used for such detection experiments to confirm these predictions. 

\begin{table}[t]
\caption{Deflection angle of UHECR in a colliding galaxy system.}
\begin{center}
\begin{tabular}{cccccc}
\hline\hline
\multicolumn{3}{c}{B-random}  && \multicolumn{2}{c}{B-regular} \\ 
\hline
$L_{BC}$   & $L_B$   & $B_{ran}$ &\,\, &  $L_B$  & $B_{reg}$  \\
kpc     & kpc   & $\mu G$   && kpc   & $\mu G$ \\ 
\hline
\multicolumn{6}{c}{galactic disk} \\

0.05     & 10    & 15       &&  3   & 10  \\ 
\multicolumn{3}{c}{$\delta=3.4\degr$} && \multicolumn{2}{c}{$\delta=16\degr$} \\ 
\hline
\multicolumn{6}{c}{bridge, tidal tail} \\
0.1      & 10    & 15       && 5   &  10  \\ 
\multicolumn{3}{c}{$\delta=4.9\degr$} && \multicolumn{2}{c}{$\delta=23\degr$} \\ 
\hline
\multicolumn{6}{c}{merger's halo} \\
2      & 200    & 0.1       && 200   & 0.01   \\ 
\multicolumn{3}{c}{$\delta=0.7\degr$} && \multicolumn{2}{c}{$\delta=1.0\degr$} \\ \hline
\end{tabular}
\end{center}
\label{t:uhecr}
{\bf Notes.} A proton of energy of $10^{20}\,$eV 
is assumed to cross regions of various sizes $L_B$, filled with magnetic field of 
different strength and coherence scale $L_{BC}$.
\end{table}

A vital issue in high-energy astrophysics is to identify sites of origin of UHECRs. There may be problems arising from deflection of UHECR 
trajectories by intergalactic magnetic fields (Yoshiguchi et al. \cite{yoshiguchi03}; Neronov \& Semikoz \cite{neronov09}; Ryu et al. \cite{ryu10}). 
Most interacting galaxies from our sample are within 100\,Mpc distance and 
their spread-out magnetic fields could in principle affect propagation of UHECR 
on their path to the Earth. To estimate a deflection angle $\delta$ between the incoming direction of UHECR
and the sky position of their sources, we follow the approach of 
Neronov \& Semikoz (\cite{neronov09}), assuming that UHECR protons with energy of 
$10^{20}$\,eV propagate through a magnetized volume of pathlength $L_B$. 
In the case of regular field, its strength $B_{reg}$ influences the deflection angle, while in the case 
of turbulent field, its strength $B_{ran}$ as well as its coherence length $L_{BC}$ have to be known. 
We divide our estimations into those for disks of interacting galaxies, for strongly 
magnetized outflows (as bridges or tidal tails), and for envelopes of merging galaxies 
filled with \ion{H}{i} gas and spread out magnetic fields.  
We take $L_{BC}=50$\,pc for galactic disks (a commonly adopted value) and the 
scaled values for the bridges and haloes (Table \ref{t:uhecr}). 

We obtained the largest deflections of $23\degr$ for the protrusions from colliding galaxies 
filled with the regular field such as in the Taffy system (Table \ref{t:uhecr}).  
In the case of galactic disks with, e.g., regular magnetic fields stretched along 
spiral arms (like in the weakly interacting NGC\,2207, Sect. \ref{s:results}) the 
deflection angle is also large ($16\degr$). The more disrupted (random) fields are 
here less important ($\delta\approx 3\degr$). 
The probability of an individual UHECR particle to cross the merger and deflect its way to the Earth 
is very low, $<10^{-6}$, taking all 300 interacting galaxies into account within 50\,Mpc 
distance, and 200\,kpc size of their magnetized haloes. 
However, large-scale shocks in colliding galaxies and centres of merging systems were 
also proposed as possible sources of UHECRs (Bhattacharjee \& Sigl \cite{bhattacharjee00}, Giller 
at al. \cite{giller03}, Biermann et al. \cite{biermann09}). 
If interacting galaxies generate some UHECRs by large-scale shocks 
or AGNs within them, their magnetic fields would severely deflect their trajectories 
(up to $\approx23\degr$). Propagation of UHECRs through the disk or magnetized outflows 
(bridges, tidal tails) would make it very difficult to properly associate observed 
UHECRs with the sites of their origin.

We note that the magnetized halo around mergers, however large in extent, would only marginally impact 
the UHECR propagation (deflection up to about $1\degr$), owing to dispersed, weak magnetic fields (Table \ref{t:uhecr}).
The propagation of protons through the Galaxy would also result in a small deflection of up to 
about $1\degr$ (for $B_{ran}\approx4\,\mu$G).

\section{Summary and conclusions}
\label{s:summary}

We studied the influence of gravitational interactions on the distribution of 
radio emission and the properties of magnetic fields in galaxies.
We perused archival radio data from the VLA at 4.86\,GHz and 1.4\,GHz for a sample 
of 24 galaxies in 16 interacting systems. Different stages 
of interaction were approximately described by the interaction stage ($IS$) parameter 
(Sect. \ref{s:discussion}). The main conclusions are as follows.

\begin{itemize}

\item The estimated total magnetic field strengths are from 5\,$\mu$G (ARP\,222) 
to 25-27\,$\mu$G (NGC\,3256, ARP\,220), with a mean of $14\pm 5\,\mu$G, greater than
for bright non-interacting galaxies. The mean field regularity of $0.27\pm0.09$ is less than 
for typical spirals, which indicates enhanced production of random magnetic fields in the interacting objects.

\item
Detailed analysis of NGC\,2207/IC\,2163 reveals a region 
of enhanced radio total and polarized emission in the eastern part of NGC\,2207, 
which can suggest a supply of CR electrons originating in IC\,2163.
Our studies of polarized properties of this system contradict the earlier
hypothesis that the enhanced radio emission in this region is due
to compression of magnetic fields (Elmegreen et al. \cite{elmegreen95a}). 

\item The structure of magnetic field $\vec{B}$-vectors in the polar-ring galaxy 
NGC\,660 resembles an X-shape pattern, typical of late-type edge-on 
galaxies, which challenges some earlier suggestions that this is a post-merger 
system.

\item
Even in weakly interacting galaxies, the polarized emission and the orientation of 
magnetic field $\vec{B}$-vectors are very sensitive tools for revealing global galactic 
distortions (Taffy galaxies) or the direction of gas flow in tidally 
stretched spiral arms and tidal tails (NGC\,2207, NCG\,6907, NGC\,4038). 

\item
We built a tentative scenario of magnetic field evolution in interacting 
galaxies. For weak interactions the strength of magnetic field is almost constant 
($10-15\,\mu$G), then it increases up to $2\times$ with advancing 
of interaction to later stages, and decreases again, down to the limiting value 
of $5-6\,\mu$G, for old post-mergers. We suggest that the main production 
of magnetic fields in colliding 
galaxies must terminate close to the nuclear coalescence, after which 
magnetic field diffuses or is kept at the level of turbulent energy of ISM. 
The revealed evolution of magnetic fields agrees with the scenario of 
formation of ellipticals from mergers (Toomre \& Toomre \cite{toomre72}), 

\item Magnetic field strength is weakly regulated by the galactic star-forming
activity (e.g. SFR), showing a wide spread of estimated strength values (with a slope 
$\alpha=0.21\pm 0.02$ and correlation $\rho=0.63$). The dependence is stronger
in galactic centres and weaker outside. There is also a hint that the regular 
field is anticorrelated with SFR ($\rho=-0.55$). 

\item 
Asymmetry in distribution of polarized emission is stronger ($A_P=0.75$) than in total 
radio intensity ($A_T=0.64$), which can indicate that the regular component of magnetic fields
is much more sensitive to morphological distortions induced by tidal interactions 
than the random magnetic field. This means that, the asymmetry in the radio polarized 
emission could be yet another indicator of an ongoing merging process.
The asymmetries visible in the radio emission do not statistically depend 
on the global or local SFRs, while  they increase (especially in the polarization) with advancing interaction.

\item  
The constructed radio-FIR relation for interacting and non-interacting 
galaxies shows that the tidally induced intense 
star formation, morphological distortions, and processes mutually transforming 
magnetic field components (e.g. by compression and shearing) still lead to producing magnetic energy and cosmic rays that are balanced 
by the production of thermal energy and dust radiation. 
The interacting galaxies constitute the group of statistically 
strongest radio and FIR emitters. 
We also argued that the configuration of ordered magnetic fields in the 
bridge of the Taffy system indicates that the modelling of 
Lisenfeld \& Voelk (\cite{lisenfeld10}), adopted to account for the strong radio emission 
in the bridge, might not be applicable here. 

\item
The process of strong gravitational interactions can efficiently magnetize the merger surroundings
(likely up to about 200\,kpc), exerting a similar impact on IGM to its magnetization by supernova explosions and galactic winds. 
However, tidal interactions are likely insufficient to fill IGM with magnetic fields with a large volume filling factor of $\approx60\%$ as recently suggested by Dolag et al. 
(\cite{dolag10}).
If interacting galaxies generate some UHECRs by large-scale shocks 
or AGNs, their propagation through the disk or magnetized outflows 
(bridges or tidal tails) can deflect them up to $23\degr$ and make 
association of observed UHECRs with the sites of their origin less obvious.

\end{itemize}

Among the interacting systems, those at the final stages of merging as well as the post-merging systems
(as elliptical galaxies) are the least studied ones with respect to magnetic field evolution. After the interaction-induced starburst declines, 
the production of cosmic ray electrons ceases and cannot further enlighten magnetic 
fields (Sect \ref{s:evolution}). Such fields may still exist in the mergers, before they disperse because of turbulent diffusivity, but could only be discovered by the Faraday rotation methods 
applied to the background sources. Thus there is a need for very high-resolution 
and sensitive radio spectropolarimetric data from interferometers, not yet available. 
Future EVLA, LOFAR and SKA instrumentation could reveal for the first time such 
fading or relic magnetic fields, if they are actually present in the post-merging systems. 

\begin{acknowledgements}
We thank an anonymous referee for helpful comments and
suggestions. This work was supported by the Polish Ministry of Science and Higher
Education, grant 3033/B/H03/2008/35. We acknowledge the use of the HyperLeda
(http://leda.univ-lyon1.fr) and NED (http://nedwww.ipac.caltech.edu) databases.
\end{acknowledgements}

\appendix
\section{Derivation of magnetic field strengths}
\label{s:formula}

The strength of the total  $B_{tot}$  and regular $B_{reg}$ magnetic field can be estimated 
using radio polarimetric observations from the total synchrotron intensity $I_n$ and its degree of 
linear polarization $P_{nth}$. After taking the equipartition between the energy 
densities of the magnetic field and cosmic rays into account, the total magnetic field 
is given by (Beck \& Krause~\cite{beck05}) 
\begin{equation}
B_{tot} = \left[\frac{4\pi(2\alpha_n+1)(K_0+1)I_{n}E_p^{1-2\alpha_n}(\frac{\nu}{2c_1})^{\alpha_n}}{(2\alpha_n-1)c_2\alpha_nLc_3}\right]^{1\over{\alpha_n+3}}
\label{e:B}
\end{equation}
where $K_0$ is the ratio of proton to electron number densities, $\alpha_n$ the 
mean synchrotron spectral index, $L$ the pathlength through the synchrotron emitting medium, 
$E_p$ the rest energy of the proton, and 
\begin{equation}
c_1 = \frac{3e}{4\pi m_e^3c^5} = 6.2648 \times 10^{18} {\rm erg}^{-2}{\rm s}^{-1}{\rm G}^{-1},
\end{equation}

\begin{equation}
c_2(\alpha_n) = {1\over4}c_3 \frac{(\alpha_n+{5\over3})}{(\alpha_n+1)} 
\Gamma\left[{(3\alpha_n+1)\over6}\right]\times\Gamma\left[{(3\alpha_n+5)\over6}\right].
\end{equation}
Constants $c_2$ and $c_3$ are geometrical corrections that depend on magnetic field orientation. 
For a region where the field is totally regular and has a constant 
inclination $i$ with respect to the sky plane, $c_3 = (\cos i)^{\alpha_n+1}$. For a completely 
turbulent field, $c_3 = (2/3)^{\alpha_n+1)/2}$. If the synchrotron intensity is averaged over a 
large volume, such as the whole galaxy, the value of $c_3$ has to be replaced by its average over 
all occurring values of $i$. For instance, the strength of the mean regular magnetic field 
in the galactic disk can be obtained from the estimated nonthermal degree of polarization 
(Segalovitz et al.~\cite{segalovitz76}):
\begin{equation}
P_{nth} = \left(\frac{3\gamma+3}{3\gamma+7}\right)\times \left[1+\frac{(1-q)\pi^{1\over2}\Gamma[(\gamma+5)/4]}{2q\Gamma[(\gamma+7)/4]F(i)}\right]^{-1},
\end{equation}
where
\begin{equation}
F(i) = {1\over2\pi}\int_0^{2\pi}(1-{\rm sin}^2i\,{\rm sin}^2\theta)^{(\gamma+1)/4}{\rm d}\theta,
\end{equation}
$q^{1/(1+\alpha_n)}=B_{reg}/B_{turb}$, $\theta$ is 
the azimuthal angle, and $B_{turb}$ the turbulent component of the magnetic field.

We assume $K_0\approx100$  in Eq. \ref{e:B}, which is consistent with acceleration of 
cosmic rays in supernova remnants and with the local Galactic cosmic ray data. For synchrotron sources, which have a volume filling factor the pathlength through the 
source $L$ has to be replaced by $L \times f$. However, the equipartition magnetic field 
strength only weakly depends on $f$ e.g. its decrease from 1 to 0.8 results in the 
increase of field strength by less than $6\%$.

\end{document}